\documentclass[prl,reprint,twocolumn,preprintnumbers,amsmath,amssymb,
footinbib,superscriptaddress]{revtex4-1}

\usepackage[titletoc,toc,title]{appendix}
\usepackage{braket}
\usepackage[final]{feynmp}
\usepackage{ifpdf}
\usepackage{comment}
\usepackage{amsmath}
\usepackage{mathrsfs}
\usepackage{color}
\usepackage{bm}
\usepackage[font=small]{subcaption}
\usepackage[font=small]{caption}

\makeatletter
\def\endfmffile{%
	\fmfcmd{\p@rcent\space the end.^^J%
			end.^^J%
			endinput;}%
	\if@fmfio
		\immediate\closeout\@outfmf
	\fi
	\ifnum\pdfshellescape=\@ne
		\immediate\write18{mpost \thefmffile}%
	\fi}
\makeatother

\usepackage[demo]{graphicx}
\usepackage{dcolumn}
\usepackage{bm}
\usepackage{hyperref}

\newcommand{\beq}{\begin{equation}}
\newcommand{\eeq}{\end{equation}}

\begin{document}


\title{Multiply Quantized Vortices in Fermionic Superfluids: Angular Momentum, Unpaired Fermions, and Spectral Asymmetry}

\author{Abhinav Prem}
\thanks{Corresponding author: {\tt abhinav.prem@colorado.edu}}
\affiliation{Department of Physics and Center for Theory of Quantum Matter, University of Colorado, Boulder, Colorado 80309, USA}
\author{Sergej Moroz}
\affiliation{Department of Physics, Technical University of Munich, 85748 Garching, Germany}
\affiliation{Kavli Institute for Theoretical Physics, University of California, Santa Barbara, Santa Barbara, California 93106, USA}
\author{Victor Gurarie}
\affiliation{Department of Physics and Center for Theory of Quantum Matter, University of Colorado, Boulder, Colorado 80309, USA}
\author{Leo Radzihovsky}
\affiliation{Department of Physics and Center for Theory of Quantum Matter, University of Colorado, Boulder, Colorado 80309, USA}
\affiliation{Kavli Institute for Theoretical Physics, University of California, Santa Barbara, Santa Barbara, California 93106, USA}

\begin{abstract}
We compute the orbital angular momentum $L_z$ of an $s$-wave paired superfluid in the presence of an axisymmetric multiply quantized vortex. For vortices with winding number $|k| > 1$, we find that in the weak-pairing BCS regime $L_z$ is significantly reduced from its value $\hbar N k/2$ in the Bose-Einstein condensation (BEC) regime, where $N$ is the total number of fermions. This deviation results from the presence of unpaired fermions in the BCS ground state, which arise as a consequence of spectral flow along the vortex sub-gap states. We support our results analytically and numerically by solving the Bogoliubov-de-Gennes equations within the weak-pairing BCS regime.
\end{abstract}

\maketitle


Quantized vortices are a hallmark of superfluids (SFs) and superconductors. These topological defects form in response to external rotation or magnetic field and play a key role in understanding a broad spectrum of phenomena, such as the Berezinskii-Kosterlitz-Thouless transition in two-dimensional (2D) SFs~\cite{Berezinskii1971, Kosterlitz1973}, superconductor/insulator transitions~\cite{Fisher1990,Peskin,DasguptaHalperin}, turbulence~\cite{Feynman1955}, and dissipation~\cite{Anderson1966, Bardeen1965}. In fermionic $s$-wave paired states, the structure of the ground state and low lying excitations of an axisymmetric singly quantized vortex has been established through analytical and numerical studies in both the strong-pairing regime (where the SF phase is understood as a Bose-Einstein condensate (BEC) of bosonic molecules) and in the weak-pairing Bardeen Cooper Schrieffer (BCS) regime. In the BEC regime, the microscopic Gross-Pitaevskii equation provides a reliable framework~\cite{Gross1961, Pitaevskii1961}, while in the BCS regime the (self-consistent) Bogoliubov-deGennes (BdG) theory is key in identifying the structure of the ground state~\cite{Nygaard2003, Sensarma2006} and the spectrum of sub-gap fermionic excitations~\cite{Caroli1964}. \vspace{-0.5pt}

Multiply quantized vortices (MQVs) have however not received much attention. Generically in a homogeneous bulk system, the logarithmic repulsion between vortices, which scales as the square of the vortex winding number $k$, energetically favors an instability of a multiply quantized vortex into separated elementary unit vortices~\cite{pethick2008}. However, MQVs are of interest since under certain circumstances, the interaction between vortices is not purely repulsive and can support multi-vortex bound states, at least as metastable defects. This can happen, for instance, in type-II mesoscopic superconductors, where MQVs have been predicted~\cite{Schweigert1998} and experimentally observed~\cite{Geim2000,Kanda2004,Grigorieva2007,Cren2011}. In addition, it has been argued that MQVs are expected to be energetically stable in multicomponent superconductors~\cite{Dao2011,Babaev2016} and in chiral $p$-wave superconductors~\cite{Garaud2015,Sauls2009}. 
In fermionic SFs, a doubly quantized vortex was predicted~\cite{Volovik1977} and observed in $^3$He-A~\cite{Blaauwgeers2000}. It has further been argued that fast rotating Fermi gases trapped in an anharmonic potential will support an MQV state~\cite{Lundh2006,Lundh2009,Howe2009}. Similar vortex states have been created in rotating BEC experiments~\cite{Engels2003,Bretin2004,Leanhardt2002,Andersen2006}.
\begin{figure}[b]
\centering
\includegraphics[width=0.5\textwidth ,height=4.3cm]{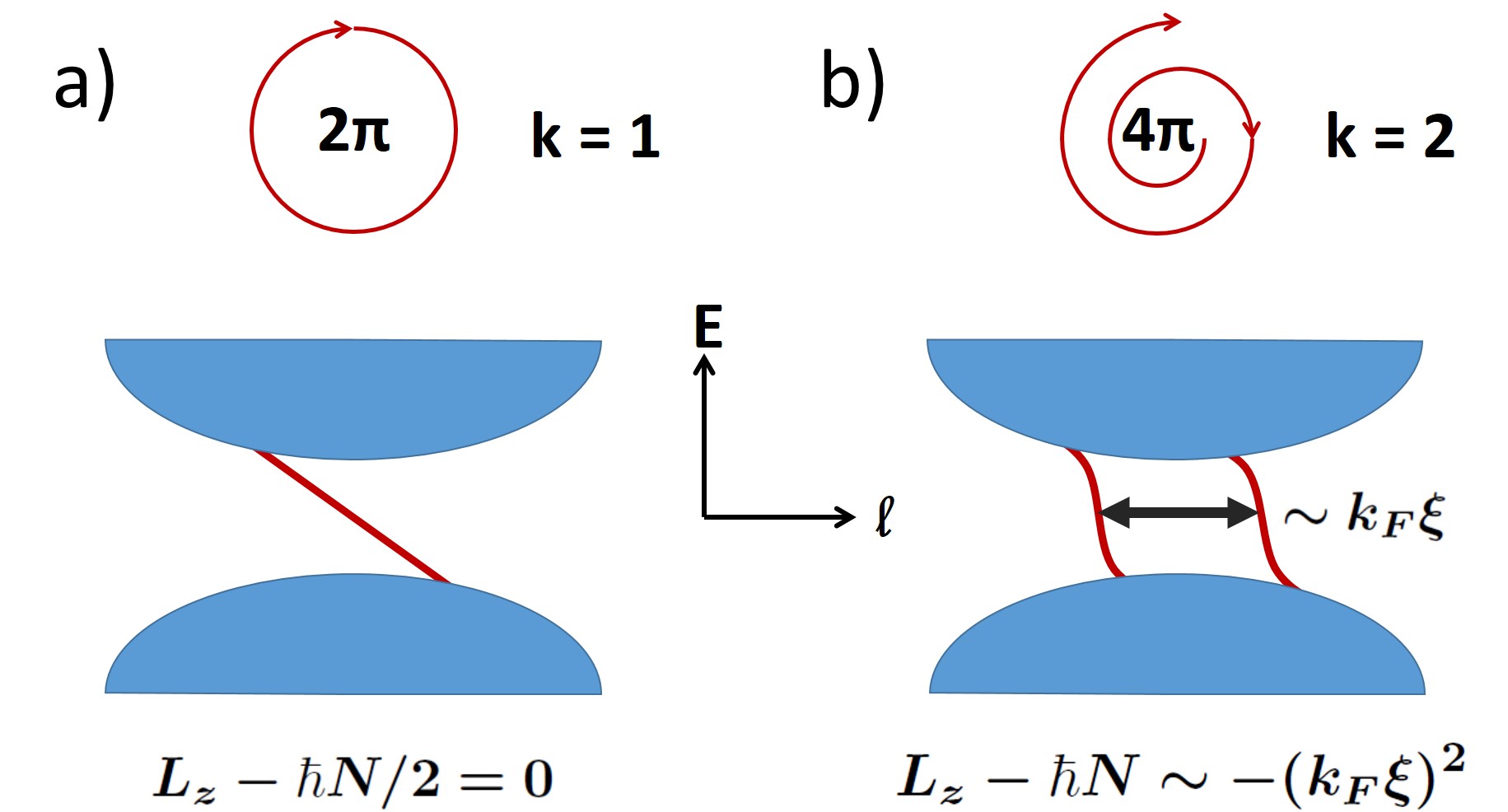}
\caption{\raggedright Summary of main result: 
a) For an elementary vortex ($k = 1$), the fermionic spectrum has a vanishing spectral asymmetry and thus all fermions are paired in the ground state, resulting in $L_z=\hbar N/2$ in the BCS regime. b) In stark contrast, for an MQV ($k=2$ pictured here as an example) mid-gap states confined to the vortex core induce a non-trivial spectral asymmetry, which leads to unpaired fermions in the ground state. These reduce $L_z$ from its na\"ive value $\hbar N$ by an amount that scales quadratically with the splitting between the red branches.}
\label{introfig}
\end{figure}

Surprisingly, as we demonstrate in this Letter, there is a fundamental difference between a singly quantized vortex ($|k|=1$) and an MQV ($|k|>1$) in a weakly-paired fermionic $s$-wave SF. This difference is manifested most clearly in the orbital angular momentum (OAM) $L_z$, as illustrated in Fig.~\ref{introfig}. 
At zero temperature in the BEC regime, a microscopic Gross-Pitaevskii calculation predicts $L_z = \hbar N k/2$, where $N$ is the total number of fermions. Intuitively, this corresponds to a simple picture where an MQV induces a quantized OAM $k$ per molecule. For an elementary vortex, this result also holds in the BCS regime, as confirmed within the self-consistent BdG framework~\cite{Nygaard2003, Sensarma2006}. As we show in this Letter, for vortices with $|k| > 1$ however, the BCS ground state contains unpaired fermions which carry OAM opposite to that carried by the Cooper pairs, thereby significantly reducing the total $L_z$ from its BEC value by an amount $\sim (k_F \xi)^2$, where $k_F$ is the Fermi momentum and $\xi$ the coherence length. While the proportionality constant is non-universal and depends on the vortex core structure, the scaling with $k_F$ and $\xi$ is robust, being independent of any boundary effects.

To derive our main result we consider a 2D\footnote{Due to the axial symmetry of the vortex line, it is sufficient to consider a two-dimensional BdG problem with a point vortex.} $s$-wave paired SF in the weak-pairing BCS regime at zero temperature within the BdG framework. 
The mean-field Hamiltonian in the presence of an axisymmetric MQV with winding number $k$ is $\hat{H} = \int d^2 r\, \Psi^\dagger \left[-\nabla^2/2 + V(r) - \mu \right] \tau_3 \Psi + \int d^2 r\, \Psi^\dagger \Delta (r) \left(e^{i k \varphi} \tau_+ + e^{-i k \varphi} \tau_- \right)\Psi$, where the Nambu spinor $\Psi = (\psi_\uparrow,\psi_\downarrow^\dagger)^T$ satisfies $\{\Psi_i(\mathbf{r}),\Psi_j^\dagger(\mathbf{r}')\} = \delta_{ij}  {\delta(\mathbf{r}-\mathbf{r}')}$. Here, $\tau_i$ are Pauli matrices, $\tau_{\pm} = (\tau_1 \pm i \tau_2)/2$, $\hbar$ and the elementary fermion mass are set to unity, and $\mu$ is the chemical potential. In principle, $\Delta(r)$ should be determined self-consistently but since our results depend only weakly on its form, we use a fixed pairing term that for our numerical analysis is taken to be $\Delta(r)$ = $\Delta_0 \tanh\left(r/\xi\right)$, where $\xi = k_F/\Delta_0$ and $\Delta_0$ is the BCS gap.

Due to the pairing term, neither the total particle number $\hat{N} = \int d^2 r\, \Psi^\dagger \tau_3 \Psi$ nor the OAM $\hat{L}_z = \int d^2 r\, \Psi^\dagger \left(-i \partial_\varphi \right) \Psi$ commutes with $\hat{H}$, and so neither are separately conserved. Instead, as pointed out in~\cite{Salomaa1987,Volovik1995}, the generalized OAM operator $\mathscr{\hat{L}} = \hat{L}_z - k\hat{N}/2$ generates a symmetry and thus, the BdG ground state and all quasi-particle excitations carry a sharp $\mathscr{\hat{L}}$ quantum number. More generally, in a chiral SF with pairing symmetry $\sim (p_x + i p_y)^\nu$ and with an MQV, the conserved operator is $\hat{L}_z - (k + \nu)\hat{N}/2$ (see Supplemental Material~\cite{supmat}). While the OAM of vortex-free chiral paired SFs ($k = 0$) was analysed in~\cite{Tada2014,Ojanen2015,Volovik2015}, here we focus on $s$-wave SFs ($\nu = 0$) with MQVs, noting that our results readily generalize to chiral states with MQVs.
 
Physically, $\mathscr{\hat{L}}$ measures the deviation of OAM in the BCS ground state from its expectation value $L_z^{\text{BEC}}= k N/2$ in the BEC regime (with $N = <\hat{N}>$). The suppression of $L_z$ in the BCS regime will hence be reflected in the eigenvalue $\mathscr{L}$ of $\mathscr{\hat{L}}$, evaluated in the ground state of the BdG Hamiltonian. We consider a disc geometry with Dirichlet boundary conditions, i.e., $V(r<R) = 0$ and $V(r>R) = \infty$. Expanding the fermionic operators in a single particle basis as $\psi_\sigma(\bm {r}) = \sum_{n,l} a_{nl\sigma} \Phi_{nl}(\bm{r})$ where $\Phi_{nl}$ satisfies $\left[-\nabla^2/2 + V(r) - \mu \right]\Phi_{nl}(\bm{r}) = \epsilon_{nl} \Phi_{nl}(\bm{r})$, the Hamiltonian becomes
\beq
\hspace{-.5cm} 
\hat{H} = \sum_{\substack{l\\n,n'}} \left(\begin{array}{c}
a_{n,l+k \uparrow}^\dagger \\
a_{n,-l \downarrow}
\end{array}
\right)^T 
\left(\begin{array}{cc}
\epsilon_{n,l+k} \delta_{n,n'} & \Delta^{(l)}_{n,n'}\\
\Delta^{(l)*}_{n,n'} & -\epsilon_{n,-l} \delta_{n,n'}
\end{array}\right) 
\left(\begin{array}{c}
a_{n',l+k \uparrow} \\
a_{n',-l \downarrow}^\dagger
\end{array}
\right)
\eeq
with $\Delta^{(l)}_{n,n'} = \int d^2 r\,\Phi^{*}_{n,l+k} \Delta(r) e^{i k \varphi} \Phi^{*}_{n',-l}$ and where $n,l$ are the radial and angular momentum quantum numbers respectively. Denoting the single-particle Hamiltonian matrix as $H^{(l)}$, particle-hole (PH) symmetry connects the different $l$-sectors through $H^{(l)*} = - C H^{(-l-k)} C^{-1}$ and the spectrum is hence PH symmetric about $l = -k/2$. 

The ground state of the BdG Hamiltonian is constructed using a generalized Bogoliubov transformation~\cite{Labonte1974, Ring} whose main steps we present here (see Supplemental Material~\cite{supmat} for details). First, we regularize the BdG Hamiltonian $H^{(l)}$ by introducing a cutoff $M\gg 1$ on $n,n'$. Generically, $H^{(l)}$ will have a different number of positive and negative eigenvalues, $M_+^{(l)}$ and $M_-^{(l)}$ respectively. The (unitary) Bogoliubov transformation is then written as
\beq
\left( \begin{array}{c}
b_{m}^{(l)}\\
d_{\bar{m}}^{(l)\dagger}
\end{array}
\right) = \sum_{n=1}^M \left(\begin{array}{cc}
S_{1,mn}^{(l)} & S_{2,mn}^{(l)}\\
S_{3,\bar{m}n}^{(l)} & S_{4,\bar{m}n}^{(l)}
\end{array}\right) \left( \begin{array}{c}
a_{n,l+k \uparrow}\\
a_{n,-l \downarrow}^\dagger
\end{array}
\right),
\eeq
where $m = 1,\dots M_+^{(l)}$, $\bar{m} = 1,\dots M_-^{(l)}$, and $M_+^{(l)}+M_-^{(l)}=2M$. The Bogoliubov operator $b_{m}^{(l)}$ annihilates a quasi-particle with positive energy $E_m^{(l)}$, $\mathscr{L}$-charge~\footnote{The quasi-particle operators carry a sharp $\mathscr{L}$-charge $l - k/2$, rather than an $l$ quantum number. Nevertheless, since the former differs from $l$ by a constant shift, it is convenient to continue labelling the states by $l$.} $l+k/2$, and spin $\uparrow$. Alternatively, by PH symmetry we can interpret it as the creation operator for a spin $\downarrow$ state with negative energy $-E_m^{(l)}$ and $\mathscr{L}$-charge $-l - k/2$. In addition, we introduce the operator $d_{\bar m}^{(l)}$ that creates a spin $\uparrow$ state with negative energy $E_{M_+^{(l)}+\bar m}^{(l)}$ and $\mathscr{L}$-charge $l+k/2$.

In terms of these operators, the ground state $\ket{BCS} \sim \otimes_l \ket{BCS}_l$ is defined as the vacuum for all positive energy quasi-particles and thus satisfies $b_{m}^{(l)}\ket{BCS} = 0$ and $d_{\bar{m}}^{(l)}\ket{BCS} = 0$. For systems with $M_+^{(l)} = M_-^{(l)}$, the ground state $\ket{BCS}$ closely resembles a Fermi sea with all negative energy states occupied
\beq
\label{wfn}
\ket{BCS} \sim \otimes_l \prod_{m=1}^{M} b_{m}^{(l)} \prod_{\bar m=1}^{M} d_{\bar m}^{(l)} \ket{0},
\eeq
where $\ket{0}$ is the Fock vacuum for $a_{n,l\sigma}$. This ground state can be understood in terms of Cooper pairs, where spin $\uparrow$ quasi-particles with $\mathscr{L}$-charge $v = l + k/2$ (created by $d^{(l)}$) are paired with quasi-particles of the opposite spin $\downarrow$ and with the opposite $\mathscr{L}$-charge $-v$ (created by $b^{(l)}$). Re-expressing the quasi-particle operators in terms of elementary fermions, we find a familiar exponential form, $\ket{BCS}_l = \exp\left(a_{n,l+k\uparrow}^\dagger K^{(l)}_{n,n'} a_{n',-l\downarrow}^\dagger\right)\ket{0}$, 
where $K^{(l)}$ is an $M \times M$ matrix (derived in Supplemental Material~\cite{supmat}), and the sum over $n,n'$ is implicit. Since $b^{(l)}$ and $d^{(l)}$ carry opposite $\mathscr{L}$-charge, the ground state Eq.~\eqref{wfn} has a vanishing $\mathscr{L}$ eigenvalue.

When $M_+^{(l)} \neq M_-^{(l)}$ however, the ground state is no longer given by Eq.~\eqref{wfn} since there will exist an imbalance between the number of quasi-particles with $\mathscr{L}$-charge $l+k/2$ and with $\mathscr{L}$-charge $-l - k/2$. This mismatch is quantified by the spectral asymmetry of the energy spectrum $\eta_l =\sum_m \text{sgn}(E_m^{(l)})= M_+^{(l)} - M_-^{(l)}$, where $\{E_m^{(l)}\}_{m\in\mathbb{N}}$ are the eigenvalues of $H^{(l)}$. In order to demonstrate that the presence of a non-trivial $\eta_l$ leads to unpaired fermions in the ground state, we perform a judiciously chosen unitary rotation on $a_{n,l\sigma}$ to a new basis of fermions $\tilde{a}_{j,l\sigma}$ via a conventional (non-Bogoliubov) rotation which does not mix creation and annihilation operators (see Supplemental Material~\cite{supmat}). Through a separate unitary rotation, we simultaneously transform the Bogoliubov operators $b^{(l)},d^{(l)}$ into a new basis $\tilde{b}^{(l)},\tilde{d}^{(l)}$. The new fermions $\tilde{a}$ and Bogoliubov quasi-particles $\tilde{b},\tilde{d}$ are related through a Bogoliubov transformation which, as always, takes the schematic form $\tilde{b} = U \tilde{a} + V \tilde{a}^\dagger$, where the matrix-valued coefficients $U,V$ satisfy $|U|^2 + |V|^2 = 1$. Following~\cite{Ring}, we find that the preceding transformations naturally distinguish between operators for which either $U$ vanishes exactly: $U=0, V = 1$ (\emph{occupied} levels), or $V$ vanishes exactly: $V = 0, U \neq 0$ (\emph{empty} levels), with the remaining operators, for which both $U,V \neq 0$, describing \emph{paired} levels. In the new basis, the ground state is superficially similar to Eq.~\eqref{wfn} since it can be expressed as 
\beq
\label{wfn2}
\ket{BCS} \sim \otimes_l \sideset{}{'}\prod_{m} \tilde{b}_{m}^{(l)} \sideset{}{'}\prod_{\bar m} \tilde{d}_{\bar m}^{(l)} \ket{0}.
\eeq
Importantly however, the restricted products here run only over paired and occupied levels. Bogoliubov operators $\tilde{b},\tilde{d}$ for empty states, which are linear super-positions of $\tilde{a}$'s, 
annihilate the bare vacuum $\ket{0}$ and are thus disallowed in Eq.~\eqref{wfn2}. Conversely, occupied states contribute to Eq.~\eqref{wfn2} but since these states create unitarily rotated fermions with certainty, $\tilde{b},\tilde{d} \sim \tilde{a}^\dagger$, they do not participate in pairing. The expression~\eqref{wfn2} is in turn equivalent to (see Supplemental Material~\cite{supmat for details}
\beq \label{gen1}
\begin{split}
\ket{BCS}_l=&\left(\prod_{i=1}^{M^{(l)}_\uparrow} \tilde a^{\dagger}_{i, l+k \uparrow} \right) \left(\prod_{i=1}^{M^{(l)}_\downarrow} \tilde a^{\dagger}_{i, -l \downarrow} \right) \\
&\times\exp\left(\sum_{j>M_\uparrow^{(l)}}^M \sum_{j' > M_\downarrow^{(l)}}^M \tilde{a}_{j,l+k\uparrow}^\dagger \mathcal{K}^{(l)}_{j,j'} \tilde{a}_{j',-l\downarrow}^\dagger \right) \ket{0},
\end{split}
\raisetag{5\baselineskip}
\eeq
where $M^{(l)}_\downarrow$ and $M^{(l)}_\uparrow$ are the number of occupied (and also empty) $\tilde{b}_m^{(l)}$ and $\tilde{d}_{\bar{m}}^{(l)}$ levels respectively. In terms of these parameters, the spectral asymmetry $\eta_l=2(M^{(l)}_\downarrow-M^{(l)}_\uparrow)$, with $M^{(l)}_{\uparrow,\downarrow} = \max(0,M - M^{(l)}_{+,-})$.

The exponential part of $\ket{BCS}$ explicitly illustrates the singlet pairing while $M_\sigma^{(l)}\neq 0$ signals the presence of unpaired fermions in the ground state.
The eigenvalue of $\mathscr{\hat L}$ can now be obtained directly from Eq.~\eqref{wfn2} by summing the individual contributions of the filled quasi-particle states and noting that $\tilde{b}^{(l)},\tilde{d}^{(l)}$ carry the same $\mathscr{L}$-charges as $b^{(l)},d^{(l)}$.
While contributions from the paired levels cancel out, the occupied levels lead to
\beq
\label{conserved}
\mathscr{L} = -\frac{1}{2}\sum_l \left(l + \frac{k}{2} \right) \eta_l.
\eeq

Alternatively, this equation can be derived directly from Eq.~\eqref{gen1}, and has previously appeared in the literature in the context of chiral SFs~\cite{Tada2014, Ojanen2015}, where $k$ is replaced by the chirality $\nu$. Physically, Eq.~\eqref{conserved} quantifies the contribution of unpaired fermions to the OAM. 

The physics originating from unpaired fermions in the ground state of a paired state was previously identified and studied in nuclear physics~\cite{Ring}, FFLO superfluids~\cite{LeoFFLO}, and chiral superfluids paired in higher partial waves~\cite{Tada2014,Volovik2015,Huang2014,Huang2015,Ojanen2015}. We now demonstrate that for a weakly-paired $s$-wave SF with an MQV, a nontrivial $\eta_l$ and the associated unpaired fermions arise as a consequence of vortex core states. 

In the BCS regime, the spectrum of the vortex core (vc) states for a singly quantized vortex $|k| = 1$ was calculated analytically by Caroli-deGennes-Matricon (CdGM)~\cite{Caroli1964} who found a single branch $E_{\text{vc}}^{(l)}$ (per spin projection) that crosses the Fermi level. This branch is PH symmetric with respect to itself, $E_{\text{vc}}^{(l)} = -E_{\text{vc}}^{(-l-1)}$ and at low energies ($E_{\text{vc}} \ll \Delta_0$) behaves linearly $E^{(l)}_{\text{vc}} = - \omega_0(l + 1/2)$, where the mini-gap $\omega_0 \sim \Delta_0/(k_F \xi)$. By numerically diagonalizing $H^{(l)}$ for $k=1$, we find that $\eta_l = 0$ for all $l$ and hence there are no unpaired fermions in the BCS ground state of an $s$-wave paired SF with an elementary vortex. Eq.~\eqref{conserved} then predicts $\mathscr{L}=0$ and thus the ground state expectation value $L_z=N/2$, which agrees with self-consistent BdG calculations~\cite{Nygaard2003}. The physics here is analogous to that of weakly paired $p+ip$ SFs, where there is a single PH symmetric edge mode that carries no OAM~\cite{Stone2004, Stone2008, Tada2014}.

\begin{figure}[t]
\centering
\begin{subfigure}[t]{0.23\textwidth}
\includegraphics[width=\textwidth]{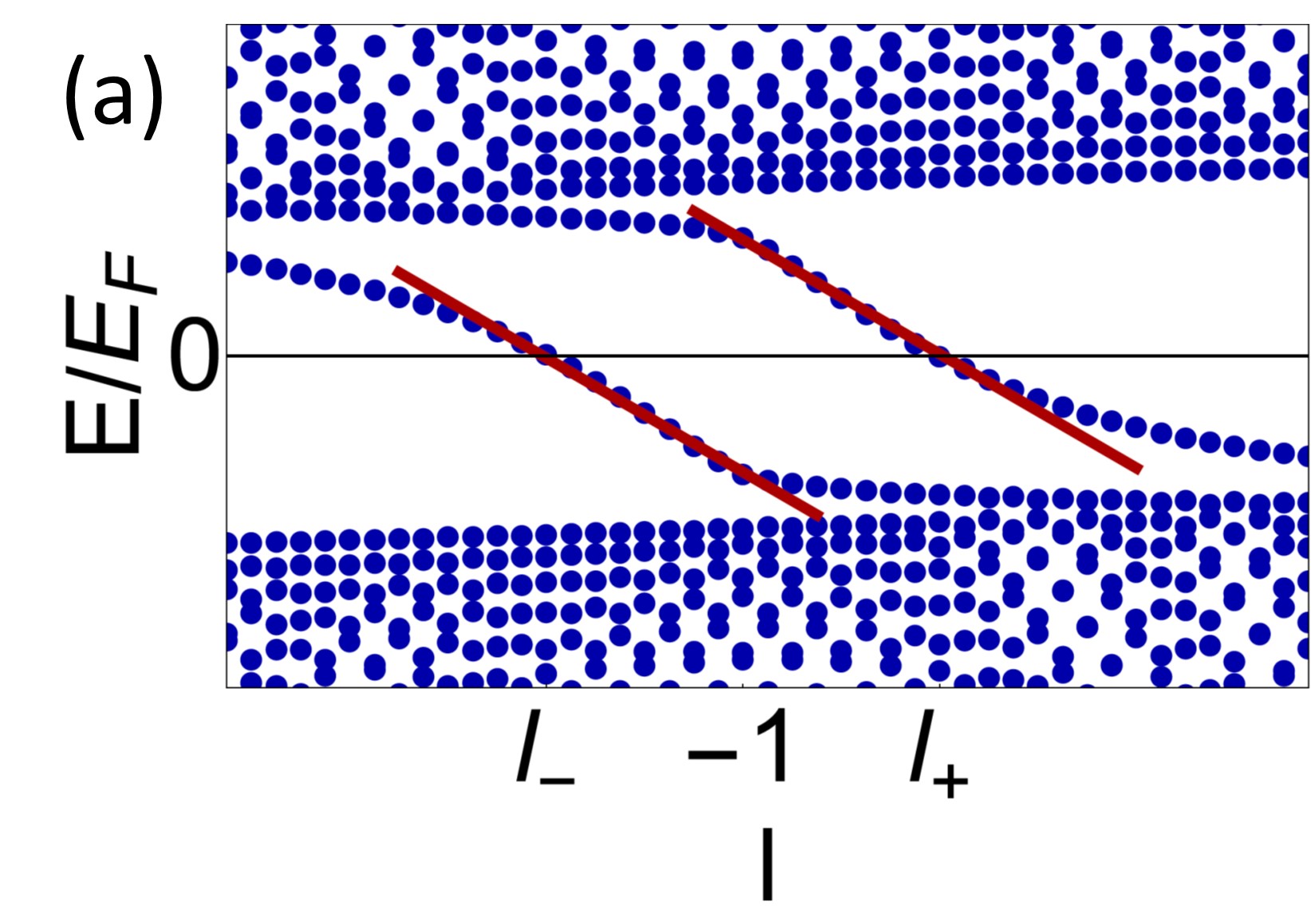}
\phantomcaption{}
\label{speccompare}
\end{subfigure}\quad
\begin{subfigure}[t]{0.23\textwidth}
\includegraphics[width=\textwidth]{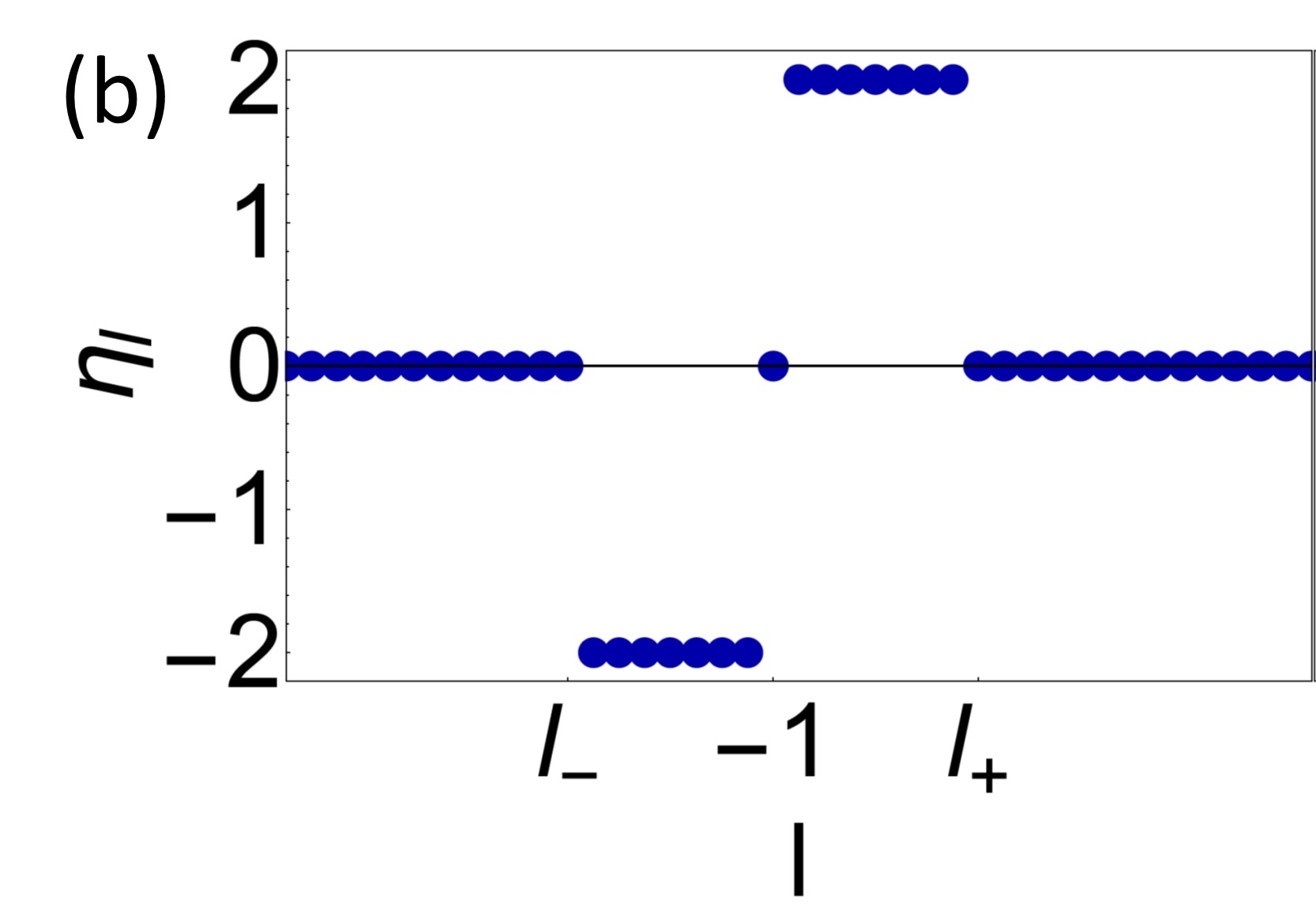}
\phantomcaption{}
\label{asymmetry}
\end{subfigure}\\ \vspace{-0.4cm}
\begin{subfigure}[t]{0.23\textwidth}
\includegraphics[width=\textwidth]{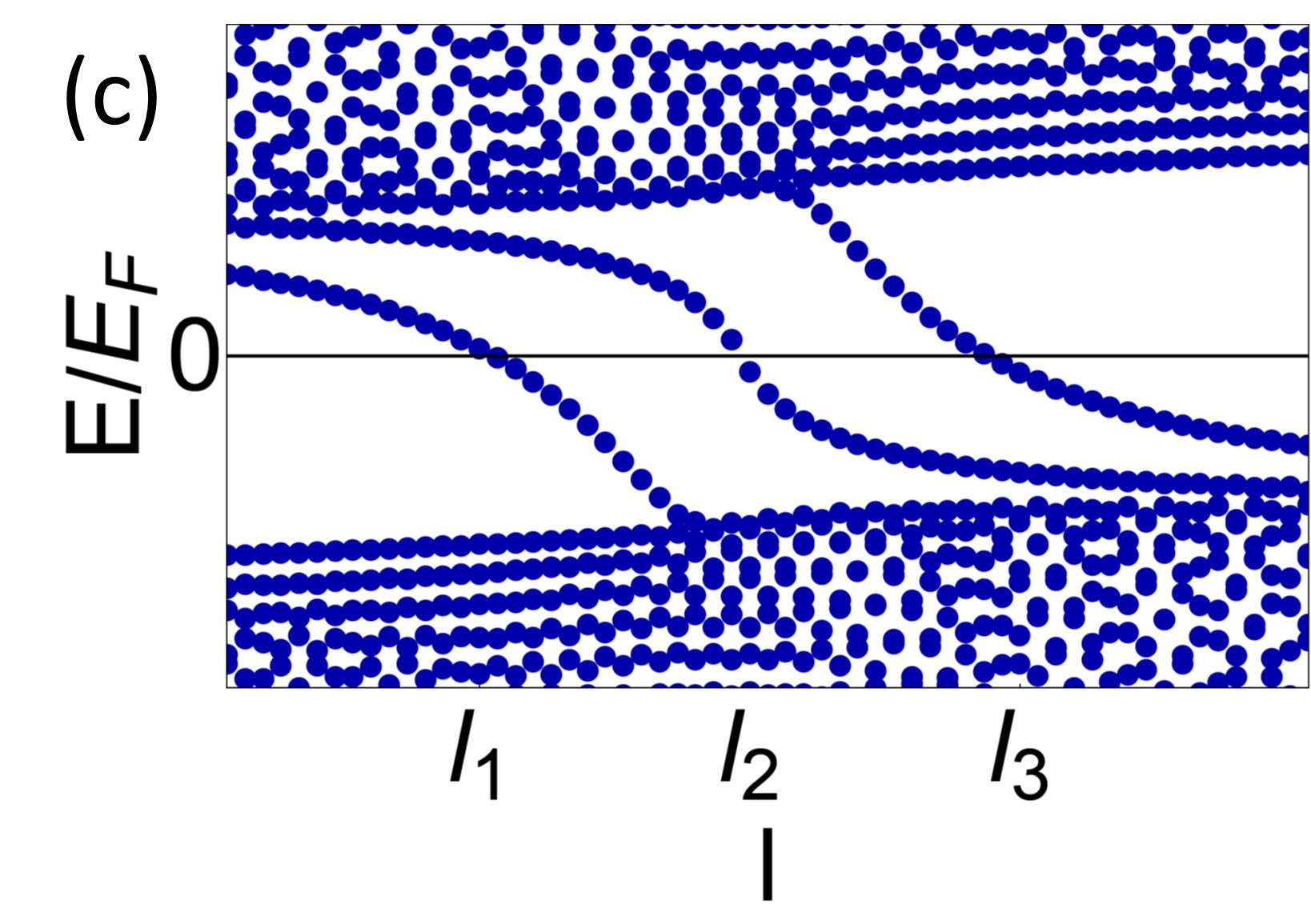}
\phantomcaption{}
\label{spec3}
\end{subfigure}\quad
\begin{subfigure}[t]{0.23\textwidth}
\includegraphics[width=\textwidth]{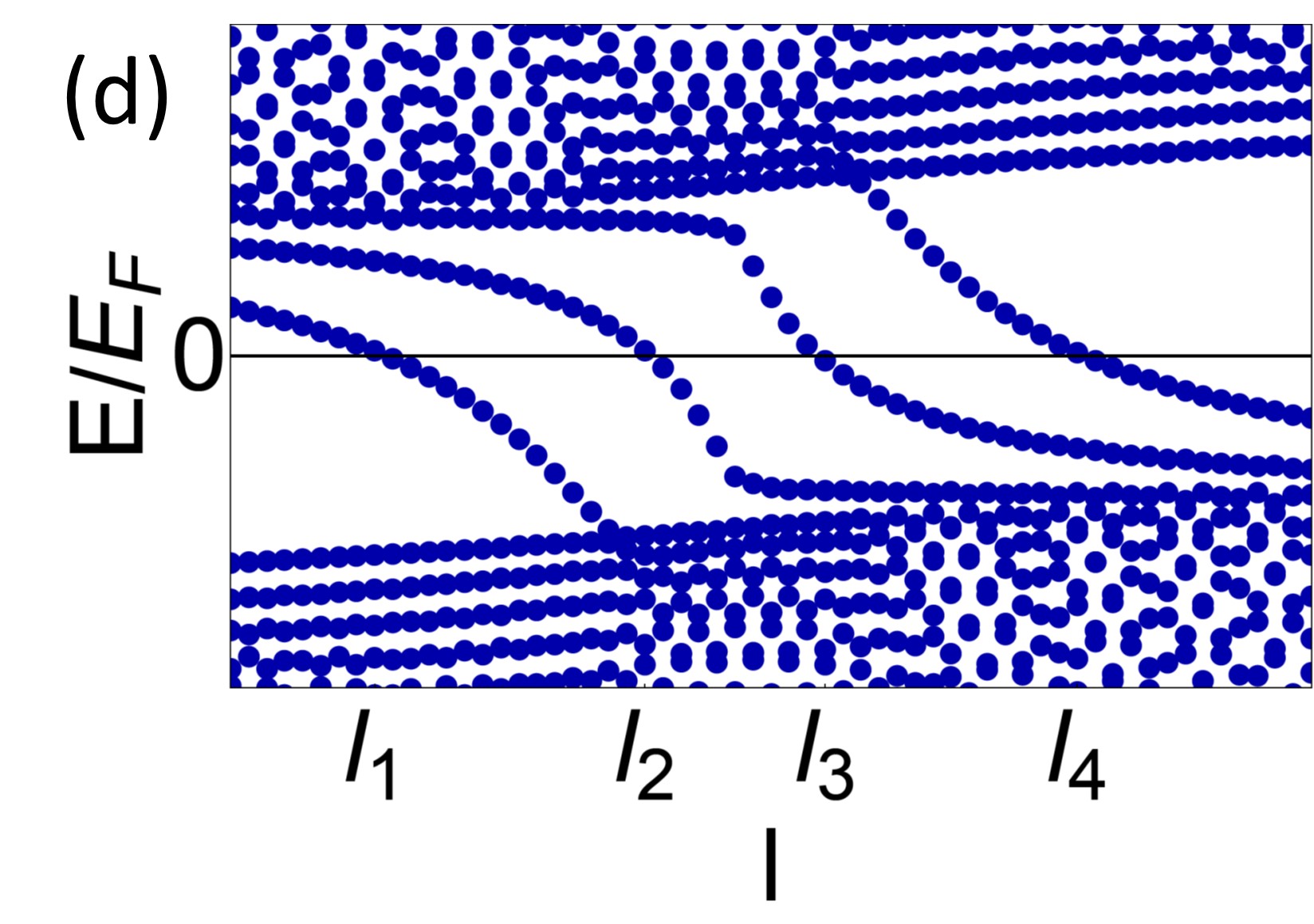}
\phantomcaption{}
\label{spec4}
\end{subfigure}
\vspace{-0.5cm}
\caption{\raggedright BdG solution for MQVs with $\Delta_0 = 0.15 E_F$, $\mu = E_F$, and $k_F R = 80$: (a) Comparison of energy spectrum for $k=2$ with analytic approximation (in red); (b) spectral asymmetry for $k=2$; (c) energy spectrum for $k=3$ and (d) for $k=4$.}
\end{figure}

For an MQV with winding number $k$, the CdGM method can be generalized and the vortex core spectrum analytically calculated within the BdG framework (see Supplemental Material~\cite{supmat}). In agreement with an argument relating the number of vortex core branches to a topological invariant~\cite{Volovik1993}, we find that $|k|$ branches (per spin projection) cross the Fermi level. At low energies these branches disperse linearly, $E_{j}(l) = -\omega_0 (l - l_j)$, where $j = 1,\dots,k$ indexes the branches and the $l_j$'s are the angular momenta at which the branches cross the Fermi level. This is consistent with results obtained by numerically diagonalizing the BdG Hamiltonian $H^{(l)}$ (for $k=2$, see Fig.~\ref{speccompare}) and with previous results on MQVs in superconductors, obtained through quasi-classical approximations~\cite{Volovik1993,Melnikov2006,Melnikov2008} and numerical simulations~\cite{Tanaka1993,Tanaka2002,Rainer1996,Salomaa1999}.

Since in the BEC regime the spectrum is completely gapped for any $k$, we find $\eta_l = 0$ for all $l$ and thus the ground state OAM is exactly $L_z = kN/2$. On the other hand, in the weakly-paired regime the energy spectrum of an MQV exhibits a nontrivial spectral asymmetry.
We consider the case $k = 2$ first (Fig.~\ref{speccompare}), where there exist two vortex core branches with linear dispersions at low energies, $E_{\text{vc},\pm}^{(l)} \sim -\omega_0 (l - l_{\pm})$ with $l_+ > l_-$. Under PH symmetry, these branches are exchanged as $E_{\text{vc},+}^{(l)} = - E_{\text{vc},-}^{(-l-2)}$ which fixes $l_- = -(l_+ + 2)$. As shown in Fig.~\ref{asymmetry}, we find that at these crossing points $\eta_l$ acquires a non-zero value: $\eta_l = -2$ for $l_-<l<-1$ and $\eta = +2$ for $-1<l<l_+$, with $\eta_l = 0$ at $l = -1$. Intuitively, this can be understood as follows---at large negative $l$, the branches are merged into the bulk and since there are no sub-gap states, $\eta_l = 0$. On increasing $l$, the branches begin separating from the bulk but since both have positive energy, $\eta_l$ still vanishes. At $l_-$ however, one of the branches crosses the Fermi energy, creating a difference of precisely two between the number of negative and positive energy eigenvalues of $H^{(l)}$. At $l = -1$, $\eta_l$ necessarily vanishes due to PH symmetry, which also fixes $\eta_l$ for $l>-1$. In contrast with $|k|= 1$, the branches are not PH symmetric with respect to themselves, allowing the spectral asymmetry to acquire a non-zero value in the BCS regime. The fact that $\eta_l$ changes from the BEC to the BCS regime can also be understood as a consequence of spectral flow along the vortex core states, since $\eta_l$ (and hence $\mathscr{L}$) cannot change its value in any other way.

A non-zero spectral asymmetry $\eta_l$ appears generally for any $|k|\geq 2$ within the BCS regime: for even $k$ (see Fig.~\ref{spec4}), there are $|k|/2$ pairs of branches such that the branches within each pair are PH symmetric with each other. $\eta_l$ then changes by $\pm 2$ whenever one of these branches crosses the Fermi level; for odd $k$ (see Fig.~\ref{spec3}) there are $(|k|-1)/2$ pairs that contribute to a non-trivial $\eta_l$, since the branches within each pair go into each other under a PH transformation, while the remaining branch is PH symmetric with respect to itself and therefore does not contribute to $\eta_l$. 

\begin{figure}[t]
\centering
\includegraphics[width=7cm]{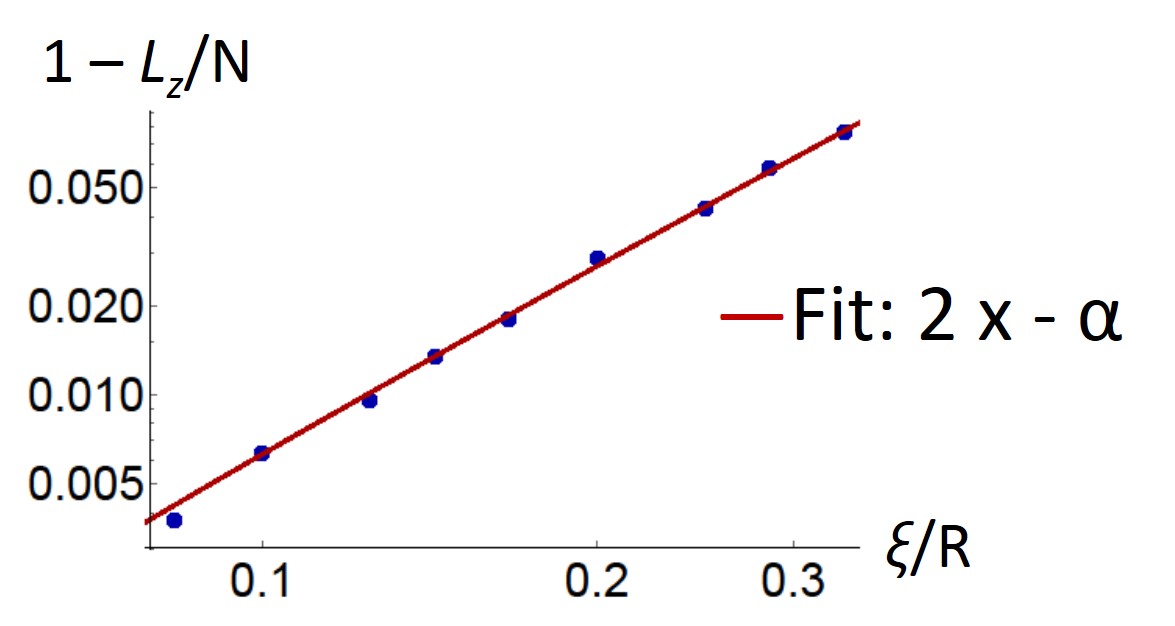}
\caption{\raggedright The analytic prediction $L_z/N = 1 - \alpha (\xi/R)^2$ (red line) fits the numerical data (blue dots) well over a wide window within the BCS regime, $0.05 \lesssim \Delta_0/E_F \lesssim 0.25$, for an MQV with $k = 2$. The slope of the fit equals two as shown on a log-log plot.}
\label{compare}
\end{figure}

Having established the existence of a non-vanishing $\eta_l$, we see that there must exist unpaired fermions in the BCS ground state for $|k| \geq 2$, and as a consequence of Eq.~\eqref{conserved}, $\mathscr{L}$ acquires a non-trivial ground state eigenvalue. For $k = 2$, this is $\mathscr{L} = -l_+^2 -l_+$, where we used PH symmetry to relate $l_-$ to $l_+$. Importantly, the analytic calculation of the vortex core states (performed in Supplemental Material~\cite{supmat}) demonstrates that the positions of the crossing points are located at $l_{\pm} \sim k_F \xi$ with the pre-factor fixed by the form of $\Delta(r)$. This scaling persists in self-consistent numerical calculations~\cite{Tanaka2002,Rainer1996,Salomaa1999}. Eq.~\eqref{conserved} along with this scaling thus establishes the reduction of the OAM of the $k=2$ MQV in the weakly paired regime. To leading order in $k_F \xi$, $\mathscr{L} = L_z - N \sim -\left(k_F \xi \right)^2$.
As a result, the OAM is significantly suppressed from $L_z^{BEC} = N$ since $k_F\xi \gg 1$ in the BCS regime ($\Delta_0 \ll E_F$). This analysis confirms that the unpaired fermions carry angular momentum opposite to that carried by the Cooper pairs.
On a disc, $N \approx (k_F R)^2/2$, leading to $L_z/N \approx 1 - \alpha \left(\xi/R\right)^2$, where $\alpha$ is an $O(1)$ constant fixed by $\Delta(r)$. As an independent check, we have verified this behavior by numerically calculating $L_z/N$ using the full BdG solution (see Supplemental Material~\cite{supmat} for details). In Fig.~\ref{compare}, the quadratic scaling is shown to be in good agreement with the numerical data. We thus expect a substantial reduction of the OAM in the BCS regime, where $\xi$ can be comparable to $R$~\cite{Geim2000}. We also expect that when two elementary vortices merge into a $k=2$ MQV~\cite{Melnikov2006,Melnikov2008}, the ground state OAM decreases from $L_z = N$ by an amount $\sim (k_F \xi)^2$.

A central feature of our result is that the suppression of $L_z$ for $|k|\geq 2$ is independent of any boundary effects and is solely determined by the splitting between the vortex core branches. Given this insensitivity to boundary details, we expect our results to hold for more general sample geometries, which may lack axial symmetry. Unlike the ground state energy, which might depend strongly on the gap profile, the OAM thus exhibits universal scaling behavior in the weak pairing BCS regime.
The lack of dependence of the OAM on the system boundary is in stark contrast with weakly-paired chiral (e.g., $d + id$) SFs, where it was shown~\cite{Tada2014,Volovik2015} that the OAM is suppressed due to the topological edge modes, but that this effect is strongly dependent on the edge details~\cite{Huang2014,Huang2015,Ojanen2015, Tada2015}. Our analysis hence suggests that $s$-wave SFs with MQVs may prove to be a more robust platform for investigating the intriguing suppression of the OAM in paired SFs. While the OAM has been measured in SFs~\cite{Chevy2000,Hodby2003,Riedl2011}, we also 
expect signatures of unpaired fermions---which create a current localized 
around the vortex core that flows counter to the superflow---in 
local supercurrent density measurements in MQV states~\cite{Kanda2004}.

\paragraph{Acknowledgements:} We acknowledge useful discussions with Egor Babaev, Masaki Oshikawa, Michael Stone, Yasuhiro Tada, and Grigory Volovik. A.P. thanks W. Cairncross for helpful comments on the draft. 
A.P. and V.G. acknowledge support by NSF grants DMR-1205303 and PHY-1211914.
The work of S.M. is supported by the Emmy Noether Programme of German Research Foundation (DFG) under grant No. MO 3013/1-1.
This research was supported
in part by the National Science Foundation under Grant
No. DMR-1001240 (L.R.), through the KITP under Grant No.
NSF PHY-1125915 (S.M. and L.R.) and by the Simons Investigator award
from the Simons Foundation (L.R.). We thank the KITP for its
hospitality during our stay as part of the ``Universality in Few-Body Systems" 
(SM), "Synthetic Quantum Matter" (L.R.), and sabbatical (L.R.) programs, when part
of this work was completed.


\bibliography{library}

\begin{thebibliography}{61}%
\makeatletter
\providecommand \@ifxundefined [1]{%
 \@ifx{#1\undefined}
}%
\providecommand \@ifnum [1]{%
 \ifnum #1\expandafter \@firstoftwo
 \else \expandafter \@secondoftwo
 \fi
}%
\providecommand \@ifx [1]{%
 \ifx #1\expandafter \@firstoftwo
 \else \expandafter \@secondoftwo
 \fi
}%
\providecommand \natexlab [1]{#1}%
\providecommand \enquote  [1]{``#1''}%
\providecommand \bibnamefont  [1]{#1}%
\providecommand \bibfnamefont [1]{#1}%
\providecommand \citenamefont [1]{#1}%
\providecommand \href@noop [0]{\@secondoftwo}%
\providecommand \href [0]{\begingroup \@sanitize@url \@href}%
\providecommand \@href[1]{\@@startlink{#1}\@@href}%
\providecommand \@@href[1]{\endgroup#1\@@endlink}%
\providecommand \@sanitize@url [0]{\catcode `\\12\catcode `\$12\catcode
  `\&12\catcode `\#12\catcode `\^12\catcode `\_12\catcode `\%12\relax}%
\providecommand \@@startlink[1]{}%
\providecommand \@@endlink[0]{}%
\providecommand \url  [0]{\begingroup\@sanitize@url \@url }%
\providecommand \@url [1]{\endgroup\@href {#1}{\urlprefix }}%
\providecommand \urlprefix  [0]{URL }%
\providecommand \Eprint [0]{\href }%
\providecommand \doibase [0]{http://dx.doi.org/}%
\providecommand \selectlanguage [0]{\@gobble}%
\providecommand \bibinfo  [0]{\@secondoftwo}%
\providecommand \bibfield  [0]{\@secondoftwo}%
\providecommand \translation [1]{[#1]}%
\providecommand \BibitemOpen [0]{}%
\providecommand \bibitemStop [0]{}%
\providecommand \bibitemNoStop [0]{.\EOS\space}%
\providecommand \EOS [0]{\spacefactor3000\relax}%
\providecommand \BibitemShut  [1]{\csname bibitem#1\endcsname}%
\let\auto@bib@innerbib\@empty
\bibitem [{\citenamefont {Berezinskii}(1970)}]{Berezinskii1971}%
  \BibitemOpen
  \bibfield  {author} {\bibinfo {author} {\bibfnamefont {V.~L.}\ \bibnamefont
  {Berezinskii}},\ }\href@noop {} {\bibfield  {journal} {\bibinfo  {journal}
  {Zh. Eksp. Teor. Fiz}\ }\textbf {\bibinfo {volume} {59}},\ \bibinfo {pages}
  {907} (\bibinfo {year} {1970})},\ \bibinfo {note} {[Sov. Phys. JETP. {\bf
  32}, 493 (1971)]}\BibitemShut {NoStop}%
\bibitem [{\citenamefont {Kosterlitz}\ and\ \citenamefont
  {Thouless}(1973)}]{Kosterlitz1973}%
  \BibitemOpen
  \bibfield  {author} {\bibinfo {author} {\bibfnamefont {J.~M.}\ \bibnamefont
  {Kosterlitz}}\ and\ \bibinfo {author} {\bibfnamefont {D.~J.}\ \bibnamefont
  {Thouless}},\ }\href {http://stacks.iop.org/0022-3719/6/i=7/a=010} {\bibfield
   {journal} {\bibinfo  {journal} {J. Phys. C: Solid St. Phys.}\ }\textbf
  {\bibinfo {volume} {6}},\ \bibinfo {pages} {1181} (\bibinfo {year}
  {1973})}\BibitemShut {NoStop}%
\bibitem [{\citenamefont {Fisher}(1990)}]{Fisher1990}%
  \BibitemOpen
  \bibfield  {author} {\bibinfo {author} {\bibfnamefont {M.~P.~A.}\
  \bibnamefont {Fisher}},\ }\href {\doibase 10.1103/PhysRevLett.65.923}
  {\bibfield  {journal} {\bibinfo  {journal} {Phys. Rev. Lett.}\ }\textbf
  {\bibinfo {volume} {65}},\ \bibinfo {pages} {923} (\bibinfo {year}
  {1990})}\BibitemShut {NoStop}%
\bibitem [{\citenamefont {Peskin}(1978)}]{Peskin}%
  \BibitemOpen
  \bibfield  {author} {\bibinfo {author} {\bibfnamefont {M.~E.}\ \bibnamefont
  {Peskin}},\ }\href {\doibase http://dx.doi.org/10.1016/0003-4916(78)90252-X}
  {\bibfield  {journal} {\bibinfo  {journal} {Ann. Phys.}\ }\textbf {\bibinfo
  {volume} {113}},\ \bibinfo {pages} {122 } (\bibinfo {year}
  {1978})}\BibitemShut {NoStop}%
\bibitem [{\citenamefont {Dasgupta}\ and\ \citenamefont
  {Halperin}(1981)}]{DasguptaHalperin}%
  \BibitemOpen
  \bibfield  {author} {\bibinfo {author} {\bibfnamefont {C.}~\bibnamefont
  {Dasgupta}}\ and\ \bibinfo {author} {\bibfnamefont {B.~I.}\ \bibnamefont
  {Halperin}},\ }\href {\doibase 10.1103/PhysRevLett.47.1556} {\bibfield
  {journal} {\bibinfo  {journal} {Phys. Rev. Lett.}\ }\textbf {\bibinfo
  {volume} {47}},\ \bibinfo {pages} {1556} (\bibinfo {year}
  {1981})}\BibitemShut {NoStop}%
\bibitem [{\citenamefont {Feynman}(1955)}]{Feynman1955}%
  \BibitemOpen
  \bibfield  {author} {\bibinfo {author} {\bibfnamefont {R.}~\bibnamefont
  {Feynman}},\ }in\ \href
  {http://www.sciencedirect.com/science/article/pii/S0079641708600773} {\emph
  {\bibinfo {booktitle} {Progress in Low Temperature Physics}}},\ Vol.~\bibinfo
  {volume} {1}\ (\bibinfo  {publisher} {Elsevier, New York},\ \bibinfo {year}
  {1955})\ pp.\ \bibinfo {pages} {17 -- 53}\BibitemShut {NoStop}%
\bibitem [{\citenamefont {Anderson}(1966)}]{Anderson1966}%
  \BibitemOpen
  \bibfield  {author} {\bibinfo {author} {\bibfnamefont {P.~W.}\ \bibnamefont
  {Anderson}},\ }\href {\doibase 10.1103/RevModPhys.38.298} {\bibfield
  {journal} {\bibinfo  {journal} {Rev. Mod. Phys.}\ }\textbf {\bibinfo {volume}
  {38}},\ \bibinfo {pages} {298} (\bibinfo {year} {1966})}\BibitemShut
  {NoStop}%
\bibitem [{\citenamefont {Bardeen}\ and\ \citenamefont
  {Stephen}(1965)}]{Bardeen1965}%
  \BibitemOpen
  \bibfield  {author} {\bibinfo {author} {\bibfnamefont {J.}~\bibnamefont
  {Bardeen}}\ and\ \bibinfo {author} {\bibfnamefont {M.~J.}\ \bibnamefont
  {Stephen}},\ }\href {\doibase 10.1103/PhysRev.140.A1197} {\bibfield
  {journal} {\bibinfo  {journal} {Phys. Rev.}\ }\textbf {\bibinfo {volume}
  {140}},\ \bibinfo {pages} {A1197} (\bibinfo {year} {1965})}\BibitemShut
  {NoStop}%
\bibitem [{\citenamefont {Gross}(1961)}]{Gross1961}%
  \BibitemOpen
  \bibfield  {author} {\bibinfo {author} {\bibfnamefont {E.~P.}\ \bibnamefont
  {Gross}},\ }\href {\doibase 10.1007/BF02731494} {\bibfield  {journal}
  {\bibinfo  {journal} {Il Nuovo Cimento (1955-1965)}\ }\textbf {\bibinfo
  {volume} {20}},\ \bibinfo {pages} {454} (\bibinfo {year} {1961})}\BibitemShut
  {NoStop}%
\bibitem [{\citenamefont {Pitaevskii}(1961)}]{Pitaevskii1961}%
  \BibitemOpen
  \bibfield  {author} {\bibinfo {author} {\bibfnamefont {L.~P.}\ \bibnamefont
  {Pitaevskii}},\ }\href@noop {} {\bibfield  {journal} {\bibinfo  {journal}
  {Sov. Phys. JETP.}\ }\textbf {\bibinfo {volume} {13}},\ \bibinfo {pages}
  {451} (\bibinfo {year} {1961})}\BibitemShut {NoStop}%
\bibitem [{\citenamefont {Nygaard}\ \emph {et~al.}(2003)\citenamefont
  {Nygaard}, \citenamefont {Bruun}, \citenamefont {Clark},\ and\ \citenamefont
  {Feder}}]{Nygaard2003}%
  \BibitemOpen
  \bibfield  {author} {\bibinfo {author} {\bibfnamefont {N.}~\bibnamefont
  {Nygaard}}, \bibinfo {author} {\bibfnamefont {G.~M.}\ \bibnamefont {Bruun}},
  \bibinfo {author} {\bibfnamefont {C.~W.}\ \bibnamefont {Clark}}, \ and\
  \bibinfo {author} {\bibfnamefont {D.~L.}\ \bibnamefont {Feder}},\ }\href
  {\doibase 10.1103/PhysRevLett.90.210402} {\bibfield  {journal} {\bibinfo
  {journal} {Phys. Rev. Lett.}\ }\textbf {\bibinfo {volume} {90}},\ \bibinfo
  {pages} {210402} (\bibinfo {year} {2003})}\BibitemShut {NoStop}%
\bibitem [{\citenamefont {Sensarma}\ \emph {et~al.}(2006)\citenamefont
  {Sensarma}, \citenamefont {Randeria},\ and\ \citenamefont
  {Ho}}]{Sensarma2006}%
  \BibitemOpen
  \bibfield  {author} {\bibinfo {author} {\bibfnamefont {R.}~\bibnamefont
  {Sensarma}}, \bibinfo {author} {\bibfnamefont {M.}~\bibnamefont {Randeria}},
  \ and\ \bibinfo {author} {\bibfnamefont {T.-L.}\ \bibnamefont {Ho}},\ }\href
  {\doibase 10.1103/PhysRevLett.96.090403} {\bibfield  {journal} {\bibinfo
  {journal} {Phys. Rev. Lett.}\ }\textbf {\bibinfo {volume} {96}},\ \bibinfo
  {pages} {090403} (\bibinfo {year} {2006})}\BibitemShut {NoStop}%
\bibitem [{\citenamefont {Caroli}\ \emph {et~al.}(1964)\citenamefont {Caroli},
  \citenamefont {{De Gennes}},\ and\ \citenamefont {Matricon}}]{Caroli1964}%
  \BibitemOpen
  \bibfield  {author} {\bibinfo {author} {\bibfnamefont {C.}~\bibnamefont
  {Caroli}}, \bibinfo {author} {\bibfnamefont {P.}~\bibnamefont {{De Gennes}}},
  \ and\ \bibinfo {author} {\bibfnamefont {J.}~\bibnamefont {Matricon}},\
  }\href {\doibase 10.1016/0031-9163(64)90375-0} {\bibfield  {journal}
  {\bibinfo  {journal} {Phys. Lett.}\ }\textbf {\bibinfo {volume} {9}},\
  \bibinfo {pages} {307} (\bibinfo {year} {1964})}\BibitemShut {NoStop}%
\bibitem [{\citenamefont {Pethick}\ and\ \citenamefont
  {Smith}(2008)}]{pethick2008}%
  \BibitemOpen
  \bibfield  {author} {\bibinfo {author} {\bibfnamefont {C.~J.}\ \bibnamefont
  {Pethick}}\ and\ \bibinfo {author} {\bibfnamefont {H.}~\bibnamefont
  {Smith}},\ }\href {\doibase 10.1017/CBO9780511802850} {\emph {\bibinfo
  {title} {Bose Einstein Condensation in Dilute Gases:}}},\ \bibinfo {edition}
  {2nd}\ ed.\ (\bibinfo  {publisher} {Cambridge University Press},\ \bibinfo
  {address} {Cambridge},\ \bibinfo {year} {2008})\BibitemShut {NoStop}%
\bibitem [{\citenamefont {Schweigert}\ \emph {et~al.}(1998)\citenamefont
  {Schweigert}, \citenamefont {Peeters},\ and\ \citenamefont
  {Deo}}]{Schweigert1998}%
  \BibitemOpen
  \bibfield  {author} {\bibinfo {author} {\bibfnamefont {V.~A.}\ \bibnamefont
  {Schweigert}}, \bibinfo {author} {\bibfnamefont {F.~M.}\ \bibnamefont
  {Peeters}}, \ and\ \bibinfo {author} {\bibfnamefont {P.~S.}\ \bibnamefont
  {Deo}},\ }\href {\doibase 10.1103/PhysRevLett.81.2783} {\bibfield  {journal}
  {\bibinfo  {journal} {Phys. Rev. Lett.}\ }\textbf {\bibinfo {volume} {81}},\
  \bibinfo {pages} {2783} (\bibinfo {year} {1998})}\BibitemShut {NoStop}%
\bibitem [{\citenamefont {Geim}\ \emph {et~al.}(2000)\citenamefont {Geim},
  \citenamefont {Dubonos}, \citenamefont {Palacios}, \citenamefont
  {Grigorieva}, \citenamefont {Henini},\ and\ \citenamefont
  {Schermer}}]{Geim2000}%
  \BibitemOpen
  \bibfield  {author} {\bibinfo {author} {\bibfnamefont {A.~K.}\ \bibnamefont
  {Geim}}, \bibinfo {author} {\bibfnamefont {S.~V.}\ \bibnamefont {Dubonos}},
  \bibinfo {author} {\bibfnamefont {J.~J.}\ \bibnamefont {Palacios}}, \bibinfo
  {author} {\bibfnamefont {I.~V.}\ \bibnamefont {Grigorieva}}, \bibinfo
  {author} {\bibfnamefont {M.}~\bibnamefont {Henini}}, \ and\ \bibinfo {author}
  {\bibfnamefont {J.~J.}\ \bibnamefont {Schermer}},\ }\href {\doibase
  10.1103/PhysRevLett.85.1528} {\bibfield  {journal} {\bibinfo  {journal}
  {Phys. Rev. Lett.}\ }\textbf {\bibinfo {volume} {85}},\ \bibinfo {pages}
  {1528} (\bibinfo {year} {2000})}\BibitemShut {NoStop}%
\bibitem [{\citenamefont {Kanda}\ \emph {et~al.}(2004)\citenamefont {Kanda},
  \citenamefont {Baelus}, \citenamefont {Peeters}, \citenamefont {Kadowaki},\
  and\ \citenamefont {Ootuka}}]{Kanda2004}%
  \BibitemOpen
  \bibfield  {author} {\bibinfo {author} {\bibfnamefont {A.}~\bibnamefont
  {Kanda}}, \bibinfo {author} {\bibfnamefont {B.~J.}\ \bibnamefont {Baelus}},
  \bibinfo {author} {\bibfnamefont {F.~M.}\ \bibnamefont {Peeters}}, \bibinfo
  {author} {\bibfnamefont {K.}~\bibnamefont {Kadowaki}}, \ and\ \bibinfo
  {author} {\bibfnamefont {Y.}~\bibnamefont {Ootuka}},\ }\href {\doibase
  10.1103/PhysRevLett.93.257002} {\bibfield  {journal} {\bibinfo  {journal}
  {Phys. Rev. Lett.}\ }\textbf {\bibinfo {volume} {93}},\ \bibinfo {pages}
  {257002} (\bibinfo {year} {2004})}\BibitemShut {NoStop}%
\bibitem [{\citenamefont {Grigorieva}\ \emph {et~al.}(2007)\citenamefont
  {Grigorieva}, \citenamefont {Escoffier}, \citenamefont {Misko}, \citenamefont
  {Baelus}, \citenamefont {Peeters}, \citenamefont {Vinnikov},\ and\
  \citenamefont {Dubonos}}]{Grigorieva2007}%
  \BibitemOpen
  \bibfield  {author} {\bibinfo {author} {\bibfnamefont {I.~V.}\ \bibnamefont
  {Grigorieva}}, \bibinfo {author} {\bibfnamefont {W.}~\bibnamefont
  {Escoffier}}, \bibinfo {author} {\bibfnamefont {V.~R.}\ \bibnamefont
  {Misko}}, \bibinfo {author} {\bibfnamefont {B.~J.}\ \bibnamefont {Baelus}},
  \bibinfo {author} {\bibfnamefont {F.~M.}\ \bibnamefont {Peeters}}, \bibinfo
  {author} {\bibfnamefont {L.~Y.}\ \bibnamefont {Vinnikov}}, \ and\ \bibinfo
  {author} {\bibfnamefont {S.~V.}\ \bibnamefont {Dubonos}},\ }\href {\doibase
  10.1103/PhysRevLett.99.147003} {\bibfield  {journal} {\bibinfo  {journal}
  {Phys. Rev. Lett.}\ }\textbf {\bibinfo {volume} {99}},\ \bibinfo {pages}
  {147003} (\bibinfo {year} {2007})}\BibitemShut {NoStop}%
\bibitem [{\citenamefont {Cren}\ \emph {et~al.}(2011)\citenamefont {Cren},
  \citenamefont {Serrier-Garcia}, \citenamefont {Debontridder},\ and\
  \citenamefont {Roditchev}}]{Cren2011}%
  \BibitemOpen
  \bibfield  {author} {\bibinfo {author} {\bibfnamefont {T.}~\bibnamefont
  {Cren}}, \bibinfo {author} {\bibfnamefont {L.}~\bibnamefont
  {Serrier-Garcia}}, \bibinfo {author} {\bibfnamefont {F.}~\bibnamefont
  {Debontridder}}, \ and\ \bibinfo {author} {\bibfnamefont {D.}~\bibnamefont
  {Roditchev}},\ }\href {\doibase 10.1103/PhysRevLett.107.097202} {\bibfield
  {journal} {\bibinfo  {journal} {Phys. Rev. Lett.}\ }\textbf {\bibinfo
  {volume} {107}},\ \bibinfo {pages} {097202} (\bibinfo {year}
  {2011})}\BibitemShut {NoStop}%
\bibitem [{\citenamefont {Dao}\ \emph {et~al.}(2011)\citenamefont {Dao},
  \citenamefont {Chibotaru}, \citenamefont {Nishio},\ and\ \citenamefont
  {Moshchalkov}}]{Dao2011}%
  \BibitemOpen
  \bibfield  {author} {\bibinfo {author} {\bibfnamefont {V.~H.}\ \bibnamefont
  {Dao}}, \bibinfo {author} {\bibfnamefont {L.~F.}\ \bibnamefont {Chibotaru}},
  \bibinfo {author} {\bibfnamefont {T.}~\bibnamefont {Nishio}}, \ and\ \bibinfo
  {author} {\bibfnamefont {V.~V.}\ \bibnamefont {Moshchalkov}},\ }\href
  {\doibase 10.1103/PhysRevB.83.020503} {\bibfield  {journal} {\bibinfo
  {journal} {Phys. Rev. B}\ }\textbf {\bibinfo {volume} {83}},\ \bibinfo
  {pages} {020503} (\bibinfo {year} {2011})}\BibitemShut {NoStop}%
\bibitem [{\citenamefont {Babaev}\ \emph {et~al.}(2017)\citenamefont {Babaev},
  \citenamefont {Carlstr\'om}, \citenamefont {Silaev},\ and\ \citenamefont
  {Speight}}]{Babaev2016}%
  \BibitemOpen
  \bibfield  {author} {\bibinfo {author} {\bibfnamefont {E.}~\bibnamefont
  {Babaev}}, \bibinfo {author} {\bibfnamefont {J.}~\bibnamefont {Carlstr\'om}},
  \bibinfo {author} {\bibfnamefont {M.}~\bibnamefont {Silaev}}, \ and\ \bibinfo
  {author} {\bibfnamefont {J.}~\bibnamefont {Speight}},\ }\href
  {https://doi.org/10.1016/j.physc.2016.08.003} {\bibfield  {journal} {\bibinfo
   {journal} {Physica (Amsterdam)}\ }\textbf {\bibinfo {volume} {533C}},\
  \bibinfo {pages} {20 } (\bibinfo {year} {2017})}\BibitemShut {NoStop}%
\bibitem [{\citenamefont {Garaud}\ and\ \citenamefont
  {Babaev}(2015)}]{Garaud2015}%
  \BibitemOpen
  \bibfield  {author} {\bibinfo {author} {\bibfnamefont {J.}~\bibnamefont
  {Garaud}}\ and\ \bibinfo {author} {\bibfnamefont {E.}~\bibnamefont
  {Babaev}},\ }\href {\doibase 10.1038/srep17540} {\bibfield  {journal}
  {\bibinfo  {journal} {Sci. Rep.}\ }\textbf {\bibinfo {volume} {5}},\ \bibinfo
  {pages} {17540} (\bibinfo {year} {2015})}\BibitemShut {NoStop}%
\bibitem [{\citenamefont {Sauls}\ and\ \citenamefont
  {Eschrig}(2009)}]{Sauls2009}%
  \BibitemOpen
  \bibfield  {author} {\bibinfo {author} {\bibfnamefont {J.~A.}\ \bibnamefont
  {Sauls}}\ and\ \bibinfo {author} {\bibfnamefont {M.}~\bibnamefont
  {Eschrig}},\ }\href {\doibase 10.1088/1367-2630/11/7/075008} {\bibfield
  {journal} {\bibinfo  {journal} {New J. Phys.}\ }\textbf {\bibinfo {volume}
  {11}},\ \bibinfo {pages} {075008} (\bibinfo {year} {2009})}\BibitemShut
  {NoStop}%
\bibitem [{\citenamefont {Volovik}\ and\ \citenamefont
  {Kopnin}(1977)}]{Volovik1977}%
  \BibitemOpen
  \bibfield  {author} {\bibinfo {author} {\bibfnamefont {G.~E.}\ \bibnamefont
  {Volovik}}\ and\ \bibinfo {author} {\bibfnamefont {N.~B.}\ \bibnamefont
  {Kopnin}},\ }\href@noop {} {\bibfield  {journal} {\bibinfo  {journal} {Pis'ma
  Zh. Eksp. Teor. Fiz.}\ }\textbf {\bibinfo {volume} {25}},\ \bibinfo {pages}
  {26} (\bibinfo {year} {1977})},\ \bibinfo {note} {[JETP Lett. {\bf 25}, 22
  (1977)]}\BibitemShut {NoStop}%
\bibitem [{\citenamefont {Blaauwgeers}\ \emph {et~al.}(2000)\citenamefont
  {Blaauwgeers}, \citenamefont {Eltsov}, \citenamefont {Krusius}, \citenamefont
  {Ruohio}, \citenamefont {Schanen},\ and\ \citenamefont
  {Volovik}}]{Blaauwgeers2000}%
  \BibitemOpen
  \bibfield  {author} {\bibinfo {author} {\bibfnamefont {R.}~\bibnamefont
  {Blaauwgeers}}, \bibinfo {author} {\bibfnamefont {V.~B.}\ \bibnamefont
  {Eltsov}}, \bibinfo {author} {\bibfnamefont {M.}~\bibnamefont {Krusius}},
  \bibinfo {author} {\bibfnamefont {J.~J.}\ \bibnamefont {Ruohio}}, \bibinfo
  {author} {\bibfnamefont {R.}~\bibnamefont {Schanen}}, \ and\ \bibinfo
  {author} {\bibfnamefont {G.~E.}\ \bibnamefont {Volovik}},\ }\href {\doibase
  10.1038/35006583} {\bibfield  {journal} {\bibinfo  {journal} {Nature}\
  }\textbf {\bibinfo {volume} {404}},\ \bibinfo {pages} {471} (\bibinfo {year}
  {2000})}\BibitemShut {NoStop}%
\bibitem [{\citenamefont {Lundh}(2006)}]{Lundh2006}%
  \BibitemOpen
  \bibfield  {author} {\bibinfo {author} {\bibfnamefont {E.}~\bibnamefont
  {Lundh}},\ }\href {\doibase 10.1088/1367-2630/8/12/304} {\bibfield  {journal}
  {\bibinfo  {journal} {New J. Phys.}\ }\textbf {\bibinfo {volume} {8}},\
  \bibinfo {pages} {304} (\bibinfo {year} {2006})}\BibitemShut {NoStop}%
\bibitem [{\citenamefont {Lundh}\ and\ \citenamefont
  {Cetoli}(2009)}]{Lundh2009}%
  \BibitemOpen
  \bibfield  {author} {\bibinfo {author} {\bibfnamefont {E.}~\bibnamefont
  {Lundh}}\ and\ \bibinfo {author} {\bibfnamefont {A.}~\bibnamefont {Cetoli}},\
  }\href {\doibase 10.1103/PhysRevA.80.023610} {\bibfield  {journal} {\bibinfo
  {journal} {Phys. Rev. A}\ }\textbf {\bibinfo {volume} {80}},\ \bibinfo
  {pages} {023610} (\bibinfo {year} {2009})}\BibitemShut {NoStop}%
\bibitem [{\citenamefont {Howe}\ \emph {et~al.}(2009)\citenamefont {Howe},
  \citenamefont {Lima},\ and\ \citenamefont {Pelster}}]{Howe2009}%
  \BibitemOpen
  \bibfield  {author} {\bibinfo {author} {\bibfnamefont {K.}~\bibnamefont
  {Howe}}, \bibinfo {author} {\bibfnamefont {A.~R.~P.}\ \bibnamefont {Lima}}, \
  and\ \bibinfo {author} {\bibfnamefont {A.}~\bibnamefont {Pelster}},\ }\href
  {\doibase 10.1140/epjd/e2009-00182-9} {\bibfield  {journal} {\bibinfo
  {journal} {Eur. Phys. Journal. D}\ }\textbf {\bibinfo {volume} {54}},\
  \bibinfo {pages} {667} (\bibinfo {year} {2009})}\BibitemShut {NoStop}%
\bibitem [{\citenamefont {Engels}\ \emph {et~al.}(2003)\citenamefont {Engels},
  \citenamefont {Coddington}, \citenamefont {Haljan}, \citenamefont
  {Schweikhard},\ and\ \citenamefont {Cornell}}]{Engels2003}%
  \BibitemOpen
  \bibfield  {author} {\bibinfo {author} {\bibfnamefont {P.}~\bibnamefont
  {Engels}}, \bibinfo {author} {\bibfnamefont {I.}~\bibnamefont {Coddington}},
  \bibinfo {author} {\bibfnamefont {P.~C.}\ \bibnamefont {Haljan}}, \bibinfo
  {author} {\bibfnamefont {V.}~\bibnamefont {Schweikhard}}, \ and\ \bibinfo
  {author} {\bibfnamefont {E.~A.}\ \bibnamefont {Cornell}},\ }\href {\doibase
  10.1103/PhysRevLett.90.170405} {\bibfield  {journal} {\bibinfo  {journal}
  {Phys. Rev. Lett.}\ }\textbf {\bibinfo {volume} {90}},\ \bibinfo {pages}
  {170405} (\bibinfo {year} {2003})}\BibitemShut {NoStop}%
\bibitem [{\citenamefont {Bretin}\ \emph {et~al.}(2004)\citenamefont {Bretin},
  \citenamefont {Stock}, \citenamefont {Seurin},\ and\ \citenamefont
  {Dalibard}}]{Bretin2004}%
  \BibitemOpen
  \bibfield  {author} {\bibinfo {author} {\bibfnamefont {V.}~\bibnamefont
  {Bretin}}, \bibinfo {author} {\bibfnamefont {S.}~\bibnamefont {Stock}},
  \bibinfo {author} {\bibfnamefont {Y.}~\bibnamefont {Seurin}}, \ and\ \bibinfo
  {author} {\bibfnamefont {J.}~\bibnamefont {Dalibard}},\ }\href {\doibase
  10.1103/PhysRevLett.92.050403} {\bibfield  {journal} {\bibinfo  {journal}
  {Physical review letters}\ }\textbf {\bibinfo {volume} {92}},\ \bibinfo
  {pages} {050403} (\bibinfo {year} {2004})}\BibitemShut {NoStop}%
\bibitem [{\citenamefont {Leanhardt}\ \emph {et~al.}(2002)\citenamefont
  {Leanhardt}, \citenamefont {G\"orlitz}, \citenamefont {Chikkatur},
  \citenamefont {Kielpinski}, \citenamefont {Shin}, \citenamefont {Pritchard},\
  and\ \citenamefont {Ketterle}}]{Leanhardt2002}%
  \BibitemOpen
  \bibfield  {author} {\bibinfo {author} {\bibfnamefont {A.~E.}\ \bibnamefont
  {Leanhardt}}, \bibinfo {author} {\bibfnamefont {A.}~\bibnamefont
  {G\"orlitz}}, \bibinfo {author} {\bibfnamefont {A.~P.}\ \bibnamefont
  {Chikkatur}}, \bibinfo {author} {\bibfnamefont {D.}~\bibnamefont
  {Kielpinski}}, \bibinfo {author} {\bibfnamefont {Y.}~\bibnamefont {Shin}},
  \bibinfo {author} {\bibfnamefont {D.~E.}\ \bibnamefont {Pritchard}}, \ and\
  \bibinfo {author} {\bibfnamefont {W.}~\bibnamefont {Ketterle}},\ }\href
  {\doibase 10.1103/PhysRevLett.89.190403} {\bibfield  {journal} {\bibinfo
  {journal} {Phys. Rev. Lett.}\ }\textbf {\bibinfo {volume} {89}},\ \bibinfo
  {pages} {190403} (\bibinfo {year} {2002})}\BibitemShut {NoStop}%
\bibitem [{\citenamefont {Andersen}\ \emph {et~al.}(2006)\citenamefont
  {Andersen}, \citenamefont {Ryu}, \citenamefont {Clad\'e}, \citenamefont
  {Natarajan}, \citenamefont {Vaziri}, \citenamefont {Helmerson},\ and\
  \citenamefont {Phillips}}]{Andersen2006}%
  \BibitemOpen
  \bibfield  {author} {\bibinfo {author} {\bibfnamefont {M.~F.}\ \bibnamefont
  {Andersen}}, \bibinfo {author} {\bibfnamefont {C.}~\bibnamefont {Ryu}},
  \bibinfo {author} {\bibfnamefont {P.}~\bibnamefont {Clad\'e}}, \bibinfo
  {author} {\bibfnamefont {V.}~\bibnamefont {Natarajan}}, \bibinfo {author}
  {\bibfnamefont {A.}~\bibnamefont {Vaziri}}, \bibinfo {author} {\bibfnamefont
  {K.}~\bibnamefont {Helmerson}}, \ and\ \bibinfo {author} {\bibfnamefont
  {W.~D.}\ \bibnamefont {Phillips}},\ }\href {\doibase
  10.1103/PhysRevLett.97.170406} {\bibfield  {journal} {\bibinfo  {journal}
  {Phys. Rev. Lett.}\ }\textbf {\bibinfo {volume} {97}},\ \bibinfo {pages}
  {170406} (\bibinfo {year} {2006})}\BibitemShut {NoStop}%
\bibitem [{Note1()}]{Note1}%
  \BibitemOpen
  \bibinfo {note} {Due to the axial symmetry of the vortex line, it is
  sufficient to consider a two-dimensional BdG problem with a point
  vortex.}\BibitemShut {Stop}%
\bibitem [{\citenamefont {Salomaa}\ and\ \citenamefont
  {Volovik}(1987)}]{Salomaa1987}%
  \BibitemOpen
  \bibfield  {author} {\bibinfo {author} {\bibfnamefont {M.~M.}\ \bibnamefont
  {Salomaa}}\ and\ \bibinfo {author} {\bibfnamefont {G.~E.}\ \bibnamefont
  {Volovik}},\ }\href {\doibase 10.1103/RevModPhys.59.533} {\bibfield
  {journal} {\bibinfo  {journal} {Rev. Mod. Phys.}\ }\textbf {\bibinfo {volume}
  {59}},\ \bibinfo {pages} {533} (\bibinfo {year} {1987})}\BibitemShut
  {NoStop}%
\bibitem [{\citenamefont {Volovik}(1995)}]{Volovik1995}%
  \BibitemOpen
  \bibfield  {author} {\bibinfo {author} {\bibfnamefont {G.~E.}\ \bibnamefont
  {Volovik}},\ }\href@noop {} {\bibfield  {journal} {\bibinfo  {journal}
  {Pis'ma Zh. Eksp. Teor. Fiz.}\ }\textbf {\bibinfo {volume} {61}},\ \bibinfo
  {pages} {935} (\bibinfo {year} {1995})},\ \bibinfo {note} {[JETP Lett. {\bf
  61}, 958 (1995)]}\BibitemShut {NoStop}%
\bibitem [{sup()}]{supmat}%
  \BibitemOpen
  \href@noop {} {\ }\bibinfo {note} {See Supplemental Material at
  \href{http://link.aps.org/supplemental/10.1103/PhysRevLett.119.067003}{http://link.aps.org/
  supplemental/10.1103/PhysRevLett.119.067003} for details on the generalized
  OAM operator, construction of the ground state, analytic solution for vortex
  subgap states, and calculating observables from BdG solutions, which includes
  Refs.~\cite{Bardeen1969,Berthod2005,Moller2011}.}\BibitemShut {Stop}%
\bibitem [{\citenamefont {Tada}\ \emph {et~al.}(2015)\citenamefont {Tada},
  \citenamefont {Nie},\ and\ \citenamefont {Oshikawa}}]{Tada2014}%
  \BibitemOpen
  \bibfield  {author} {\bibinfo {author} {\bibfnamefont {Y.}~\bibnamefont
  {Tada}}, \bibinfo {author} {\bibfnamefont {W.}~\bibnamefont {Nie}}, \ and\
  \bibinfo {author} {\bibfnamefont {M.}~\bibnamefont {Oshikawa}},\ }\href
  {\doibase 10.1103/PhysRevLett.114.195301} {\bibfield  {journal} {\bibinfo
  {journal} {Phys. Rev. Lett.}\ }\textbf {\bibinfo {volume} {114}},\ \bibinfo
  {pages} {195301} (\bibinfo {year} {2015})}\BibitemShut {NoStop}%
\bibitem [{\citenamefont {Ojanen}(2016)}]{Ojanen2015}%
  \BibitemOpen
  \bibfield  {author} {\bibinfo {author} {\bibfnamefont {T.}~\bibnamefont
  {Ojanen}},\ }\href {\doibase 10.1103/PhysRevB.93.174505} {\bibfield
  {journal} {\bibinfo  {journal} {Phys. Rev. B}\ }\textbf {\bibinfo {volume}
  {93}},\ \bibinfo {pages} {174505} (\bibinfo {year} {2016})}\BibitemShut
  {NoStop}%
\bibitem [{\citenamefont {Volovik}(2015)}]{Volovik2015}%
  \BibitemOpen
  \bibfield  {author} {\bibinfo {author} {\bibfnamefont {G.~E.}\ \bibnamefont
  {Volovik}},\ }\href {\doibase 10.1134/S0021364014230155} {\bibfield
  {journal} {\bibinfo  {journal} {JETP Letters}\ }\textbf {\bibinfo {volume}
  {100}},\ \bibinfo {pages} {742} (\bibinfo {year} {2015})}\BibitemShut
  {NoStop}%
\bibitem [{\citenamefont {Labont\'e}(1974)}]{Labonte1974}%
  \BibitemOpen
  \bibfield  {author} {\bibinfo {author} {\bibfnamefont {G.}~\bibnamefont
  {Labont\'e}},\ }\href {http://projecteuclid.org/euclid.cmp/1103859661}
  {\bibfield  {journal} {\bibinfo  {journal} {Comm. Math. Phys.}\ }\textbf
  {\bibinfo {volume} {36}},\ \bibinfo {pages} {59} (\bibinfo {year}
  {1974})}\BibitemShut {NoStop}%
\bibitem [{\citenamefont {Ring}\ and\ \citenamefont {Schuck}(1980)}]{Ring}%
  \BibitemOpen
  \bibfield  {author} {\bibinfo {author} {\bibfnamefont {P.}~\bibnamefont
  {Ring}}\ and\ \bibinfo {author} {\bibfnamefont {P.}~\bibnamefont {Schuck}},\
  }\href@noop {} {\emph {\bibinfo {title} {The Nuclear Many-Body Problem}}}\
  (\bibinfo  {publisher} {Springer-Verlag},\ \bibinfo {address} {Berlin New
  York},\ \bibinfo {year} {1980})\BibitemShut {NoStop}%
\bibitem [{Note2()}]{Note2}%
  \BibitemOpen
  \bibinfo {note} {The quasi-particle operators carry a sharp $\protect
  \mathscr {L}$-charge $l - k/2$, rather than an $l$ quantum number.
  Nevertheless, since the former differs from $l$ by a constant shift, it is
  convenient to continue labelling the states by $l$.}\BibitemShut {Stop}%
\bibitem [{\citenamefont {Sheehy}\ and\ \citenamefont
  {Radzihovsky}(2007)}]{LeoFFLO}%
  \BibitemOpen
  \bibfield  {author} {\bibinfo {author} {\bibfnamefont {D.~E.}\ \bibnamefont
  {Sheehy}}\ and\ \bibinfo {author} {\bibfnamefont {L.}~\bibnamefont
  {Radzihovsky}},\ }\href {\doibase
  http://dx.doi.org/10.1016/j.aop.2006.09.009} {\bibfield  {journal} {\bibinfo
  {journal} {Ann. Phys.}\ }\textbf {\bibinfo {volume} {322}},\ \bibinfo {pages}
  {1790 } (\bibinfo {year} {2007})}\BibitemShut {NoStop}%
\bibitem [{\citenamefont {Huang}\ \emph {et~al.}(2014)\citenamefont {Huang},
  \citenamefont {Taylor},\ and\ \citenamefont {Kallin}}]{Huang2014}%
  \BibitemOpen
  \bibfield  {author} {\bibinfo {author} {\bibfnamefont {W.}~\bibnamefont
  {Huang}}, \bibinfo {author} {\bibfnamefont {E.}~\bibnamefont {Taylor}}, \
  and\ \bibinfo {author} {\bibfnamefont {C.}~\bibnamefont {Kallin}},\ }\href
  {\doibase 10.1103/PhysRevB.90.224519} {\bibfield  {journal} {\bibinfo
  {journal} {Phys. Rev. B}\ }\textbf {\bibinfo {volume} {90}},\ \bibinfo
  {pages} {224519} (\bibinfo {year} {2014})}\BibitemShut {NoStop}%
\bibitem [{\citenamefont {Huang}\ \emph {et~al.}(2015)\citenamefont {Huang},
  \citenamefont {Lederer}, \citenamefont {Taylor},\ and\ \citenamefont
  {Kallin}}]{Huang2015}%
  \BibitemOpen
  \bibfield  {author} {\bibinfo {author} {\bibfnamefont {W.}~\bibnamefont
  {Huang}}, \bibinfo {author} {\bibfnamefont {S.}~\bibnamefont {Lederer}},
  \bibinfo {author} {\bibfnamefont {E.}~\bibnamefont {Taylor}}, \ and\ \bibinfo
  {author} {\bibfnamefont {C.}~\bibnamefont {Kallin}},\ }\href {\doibase
  10.1103/PhysRevB.91.094507} {\bibfield  {journal} {\bibinfo  {journal} {Phys.
  Rev. B}\ }\textbf {\bibinfo {volume} {91}},\ \bibinfo {pages} {094507}
  (\bibinfo {year} {2015})}\BibitemShut {NoStop}%
\bibitem [{\citenamefont {Stone}\ and\ \citenamefont {Roy}(2004)}]{Stone2004}%
  \BibitemOpen
  \bibfield  {author} {\bibinfo {author} {\bibfnamefont {M.}~\bibnamefont
  {Stone}}\ and\ \bibinfo {author} {\bibfnamefont {R.}~\bibnamefont {Roy}},\
  }\href {\doibase 10.1103/PhysRevB.69.184511} {\bibfield  {journal} {\bibinfo
  {journal} {Phys. Rev. B}\ }\textbf {\bibinfo {volume} {69}},\ \bibinfo
  {pages} {184511} (\bibinfo {year} {2004})}\BibitemShut {NoStop}%
\bibitem [{\citenamefont {Stone}\ and\ \citenamefont
  {Anduaga}(2008)}]{Stone2008}%
  \BibitemOpen
  \bibfield  {author} {\bibinfo {author} {\bibfnamefont {M.}~\bibnamefont
  {Stone}}\ and\ \bibinfo {author} {\bibfnamefont {I.}~\bibnamefont
  {Anduaga}},\ }\href {\doibase 10.1016/j.aop.2007.04.020} {\bibfield
  {journal} {\bibinfo  {journal} {Ann. Phys.}\ }\textbf {\bibinfo {volume}
  {323}},\ \bibinfo {pages} {2} (\bibinfo {year} {2008})}\BibitemShut {NoStop}%
\bibitem [{\citenamefont {Volovik}(1993)}]{Volovik1993}%
  \BibitemOpen
  \bibfield  {author} {\bibinfo {author} {\bibfnamefont {G.~E.}\ \bibnamefont
  {Volovik}},\ }\href@noop {} {\bibfield  {journal} {\bibinfo  {journal}
  {Pis'ma Zh. Eksp. Teor. Fiz.}\ }\textbf {\bibinfo {volume} {57}},\ \bibinfo
  {pages} {233} (\bibinfo {year} {1993})},\ \bibinfo {note} {[JETP Lett., {\bf
  57}, 244 (1993)]}\BibitemShut {NoStop}%
\bibitem [{\citenamefont {Mel'nikov}\ and\ \citenamefont
  {Silaev}(2006)}]{Melnikov2006}%
  \BibitemOpen
  \bibfield  {author} {\bibinfo {author} {\bibfnamefont {A.~S.}\ \bibnamefont
  {Mel'nikov}}\ and\ \bibinfo {author} {\bibfnamefont {M.~A.}\ \bibnamefont
  {Silaev}},\ }\href {\doibase 10.1134/S0021364006120113} {\bibfield  {journal}
  {\bibinfo  {journal} {JETP Letters}\ }\textbf {\bibinfo {volume} {83}},\
  \bibinfo {pages} {578} (\bibinfo {year} {2006})}\BibitemShut {NoStop}%
\bibitem [{\citenamefont {Mel'nikov}\ \emph {et~al.}(2008)\citenamefont
  {Mel'nikov}, \citenamefont {Ryzhov},\ and\ \citenamefont
  {Silaev}}]{Melnikov2008}%
  \BibitemOpen
  \bibfield  {author} {\bibinfo {author} {\bibfnamefont {A.~S.}\ \bibnamefont
  {Mel'nikov}}, \bibinfo {author} {\bibfnamefont {D.~A.}\ \bibnamefont
  {Ryzhov}}, \ and\ \bibinfo {author} {\bibfnamefont {M.~A.}\ \bibnamefont
  {Silaev}},\ }\href {\doibase 10.1103/PhysRevB.78.064513} {\bibfield
  {journal} {\bibinfo  {journal} {Phys. Rev. B}\ }\textbf {\bibinfo {volume}
  {78}},\ \bibinfo {pages} {064513} (\bibinfo {year} {2008})}\BibitemShut
  {NoStop}%
\bibitem [{\citenamefont {Tanaka}\ \emph {et~al.}(1993)\citenamefont {Tanaka},
  \citenamefont {Hasegawa},\ and\ \citenamefont {Takayanagi}}]{Tanaka1993}%
  \BibitemOpen
  \bibfield  {author} {\bibinfo {author} {\bibfnamefont {Y.}~\bibnamefont
  {Tanaka}}, \bibinfo {author} {\bibfnamefont {A.}~\bibnamefont {Hasegawa}}, \
  and\ \bibinfo {author} {\bibfnamefont {H.}~\bibnamefont {Takayanagi}},\
  }\href {http://www.sciencedirect.com/science/article/pii/003810989390024H}
  {\bibfield  {journal} {\bibinfo  {journal} {Solid State Commun.}\ }\textbf
  {\bibinfo {volume} {85}},\ \bibinfo {pages} {321 } (\bibinfo {year}
  {1993})}\BibitemShut {NoStop}%
\bibitem [{\citenamefont {Tanaka}\ \emph {et~al.}(2002)\citenamefont {Tanaka},
  \citenamefont {Robel},\ and\ \citenamefont {Jank{\'{o}}}}]{Tanaka2002}%
  \BibitemOpen
  \bibfield  {author} {\bibinfo {author} {\bibfnamefont {K.}~\bibnamefont
  {Tanaka}}, \bibinfo {author} {\bibfnamefont {I.}~\bibnamefont {Robel}}, \
  and\ \bibinfo {author} {\bibfnamefont {B.}~\bibnamefont {Jank{\'{o}}}},\
  }\href {\doibase 10.1073/pnas.082096799} {\bibfield  {journal} {\bibinfo
  {journal} {PNAS}\ }\textbf {\bibinfo {volume} {99}},\ \bibinfo {pages} {5233}
  (\bibinfo {year} {2002})}\BibitemShut {NoStop}%
\bibitem [{\citenamefont {Rainer}\ \emph {et~al.}(1996)\citenamefont {Rainer},
  \citenamefont {Sauls},\ and\ \citenamefont {Waxman}}]{Rainer1996}%
  \BibitemOpen
  \bibfield  {author} {\bibinfo {author} {\bibfnamefont {D.}~\bibnamefont
  {Rainer}}, \bibinfo {author} {\bibfnamefont {J.~A.}\ \bibnamefont {Sauls}}, \
  and\ \bibinfo {author} {\bibfnamefont {D.}~\bibnamefont {Waxman}},\ }\href
  {\doibase 10.1103/PhysRevB.54.10094} {\bibfield  {journal} {\bibinfo
  {journal} {Phys. Rev. B}\ }\textbf {\bibinfo {volume} {54}},\ \bibinfo
  {pages} {10094} (\bibinfo {year} {1996})}\BibitemShut {NoStop}%
\bibitem [{\citenamefont {Virtanen}\ and\ \citenamefont
  {Salomaa}(1999)}]{Salomaa1999}%
  \BibitemOpen
  \bibfield  {author} {\bibinfo {author} {\bibfnamefont {S.~M.~M.}\
  \bibnamefont {Virtanen}}\ and\ \bibinfo {author} {\bibfnamefont {M.~M.}\
  \bibnamefont {Salomaa}},\ }\href {\doibase 10.1103/PhysRevB.60.14581}
  {\bibfield  {journal} {\bibinfo  {journal} {Phys. Rev. B}\ }\textbf {\bibinfo
  {volume} {60}},\ \bibinfo {pages} {14581} (\bibinfo {year}
  {1999})}\BibitemShut {NoStop}%
\bibitem [{\citenamefont {Tada}(2015)}]{Tada2015}%
  \BibitemOpen
  \bibfield  {author} {\bibinfo {author} {\bibfnamefont {Y.}~\bibnamefont
  {Tada}},\ }\href {\doibase 10.1103/PhysRevB.92.104502} {\bibfield  {journal}
  {\bibinfo  {journal} {Phys. Rev. B}\ }\textbf {\bibinfo {volume} {92}},\
  \bibinfo {pages} {104502} (\bibinfo {year} {2015})}\BibitemShut {NoStop}%
\bibitem [{\citenamefont {Chevy}\ \emph {et~al.}(2000)\citenamefont {Chevy},
  \citenamefont {Madison},\ and\ \citenamefont {Dalibard}}]{Chevy2000}%
  \BibitemOpen
  \bibfield  {author} {\bibinfo {author} {\bibfnamefont {F.}~\bibnamefont
  {Chevy}}, \bibinfo {author} {\bibfnamefont {K.~W.}\ \bibnamefont {Madison}},
  \ and\ \bibinfo {author} {\bibfnamefont {J.}~\bibnamefont {Dalibard}},\
  }\href {\doibase 10.1103/PhysRevLett.85.2223} {\bibfield  {journal} {\bibinfo
   {journal} {Phys. Rev. Lett.}\ }\textbf {\bibinfo {volume} {85}},\ \bibinfo
  {pages} {2223} (\bibinfo {year} {2000})}\BibitemShut {NoStop}%
\bibitem [{\citenamefont {Hodby}\ \emph {et~al.}(2003)\citenamefont {Hodby},
  \citenamefont {Hopkins}, \citenamefont {Hechenblaikner}, \citenamefont
  {Smith},\ and\ \citenamefont {Foot}}]{Hodby2003}%
  \BibitemOpen
  \bibfield  {author} {\bibinfo {author} {\bibfnamefont {E.}~\bibnamefont
  {Hodby}}, \bibinfo {author} {\bibfnamefont {S.~A.}\ \bibnamefont {Hopkins}},
  \bibinfo {author} {\bibfnamefont {G.}~\bibnamefont {Hechenblaikner}},
  \bibinfo {author} {\bibfnamefont {N.~L.}\ \bibnamefont {Smith}}, \ and\
  \bibinfo {author} {\bibfnamefont {C.~J.}\ \bibnamefont {Foot}},\ }\href
  {\doibase 10.1103/PhysRevLett.91.090403} {\bibfield  {journal} {\bibinfo
  {journal} {Phys. Rev. Lett.}\ }\textbf {\bibinfo {volume} {91}},\ \bibinfo
  {pages} {090403} (\bibinfo {year} {2003})}\BibitemShut {NoStop}%
\bibitem [{\citenamefont {Riedl}\ \emph {et~al.}(2011)\citenamefont {Riedl},
  \citenamefont {Guajardo}, \citenamefont {Kohstall}, \citenamefont
  {Denschlag},\ and\ \citenamefont {Grimm}}]{Riedl2011}%
  \BibitemOpen
  \bibfield  {author} {\bibinfo {author} {\bibfnamefont {S.}~\bibnamefont
  {Riedl}}, \bibinfo {author} {\bibfnamefont {E.~R.~S.}\ \bibnamefont
  {Guajardo}}, \bibinfo {author} {\bibfnamefont {C.}~\bibnamefont {Kohstall}},
  \bibinfo {author} {\bibfnamefont {J.~H.}\ \bibnamefont {Denschlag}}, \ and\
  \bibinfo {author} {\bibfnamefont {R.}~\bibnamefont {Grimm}},\ }\href
  {http://stacks.iop.org/1367-2630/13/i=3/a=035003} {\bibfield  {journal}
  {\bibinfo  {journal} {New Journal of Physics}\ }\textbf {\bibinfo {volume}
  {13}},\ \bibinfo {pages} {035003} (\bibinfo {year} {2011})}\BibitemShut
  {NoStop}%
\bibitem [{\citenamefont {Bardeen}\ \emph {et~al.}(1969)\citenamefont
  {Bardeen}, \citenamefont {K\"ummel}, \citenamefont {Jacobs},\ and\
  \citenamefont {Tewordt}}]{Bardeen1969}%
  \BibitemOpen
  \bibfield  {author} {\bibinfo {author} {\bibfnamefont {J.}~\bibnamefont
  {Bardeen}}, \bibinfo {author} {\bibfnamefont {R.}~\bibnamefont {K\"ummel}},
  \bibinfo {author} {\bibfnamefont {A.~E.}\ \bibnamefont {Jacobs}}, \ and\
  \bibinfo {author} {\bibfnamefont {L.}~\bibnamefont {Tewordt}},\ }\href
  {\doibase 10.1103/PhysRev.187.556} {\bibfield  {journal} {\bibinfo  {journal}
  {Phys. Rev.}\ }\textbf {\bibinfo {volume} {187}},\ \bibinfo {pages} {556}
  (\bibinfo {year} {1969})}\BibitemShut {NoStop}%
\bibitem [{\citenamefont {Berthod}(2005)}]{Berthod2005}%
  \BibitemOpen
  \bibfield  {author} {\bibinfo {author} {\bibfnamefont {C.}~\bibnamefont
  {Berthod}},\ }\href {\doibase 10.1103/PhysRevB.71.134513} {\bibfield
  {journal} {\bibinfo  {journal} {Phys. Rev. B}\ }\textbf {\bibinfo {volume}
  {71}},\ \bibinfo {pages} {134513} (\bibinfo {year} {2005})}\BibitemShut
  {NoStop}%
\bibitem [{\citenamefont {M\"oller}\ \emph {et~al.}(2011)\citenamefont
  {M\"oller}, \citenamefont {Cooper},\ and\ \citenamefont
  {Gurarie}}]{Moller2011}%
  \BibitemOpen
  \bibfield  {author} {\bibinfo {author} {\bibfnamefont {G.}~\bibnamefont
  {M\"oller}}, \bibinfo {author} {\bibfnamefont {N.~R.}\ \bibnamefont
  {Cooper}}, \ and\ \bibinfo {author} {\bibfnamefont {V.}~\bibnamefont
  {Gurarie}},\ }\href {\doibase 10.1103/PhysRevB.83.014513} {\bibfield
  {journal} {\bibinfo  {journal} {Phys. Rev. B}\ }\textbf {\bibinfo {volume}
  {83}},\ \bibinfo {pages} {014513} (\bibinfo {year} {2011})}\BibitemShut
  {NoStop}%
\end{thebibliography}%


\renewcommand{\bibnumfmt}[1]{[S#1]}
\renewcommand{\theequation}{S\arabic{equation}}
\renewcommand{\thefigure}{S\arabic{figure}}
\renewcommand{\thetable}{S.\Roman{table}}

\setcounter{equation}{0}
\setcounter{figure}{0}
\setcounter{table}{0}

\clearpage
\onecolumngrid


\section{Supplemental Material}
\onecolumngrid



\subsection{A. Proof of $[\mathscr{\hat{L}},\hat{H}] = 0$}

Consider the BdG Hamiltonian for a two-dimensional superfluid (SF) with an MQV carrying the winding number $k$ 
\beq
H = \int d^2r\,\sum_{\sigma = \uparrow,\downarrow}\psi_{\sigma}^\dagger \left(-\frac{\nabla^2}{2} + V(r) - \mu \right)\psi_{\sigma} + \big[ \int d^2r\, \psi_{\uparrow}^{\dagger}\Delta(r) e^{i k \varphi/2}(p_x + i p_y)^\nu  e^{i k \varphi/2} \psi_{\downarrow}^{\dagger} + \text{h.c.} \big]
\eeq
Here the case $\nu=0$ corresponds to the $s$-wave spin-singlet paring, while spin-singlet chiral SFs are obtained by taking non-zero even values of $\nu$.
The fermionic operators satisfy canonical anti-commutation relations: $\{\psi_{\sigma}(\mathbf{r}),\psi_{\sigma'}^{\dagger}(\mathbf{r}')\} = \delta_{\sigma\sigma'}\delta(\mathbf{r} - \mathbf{r'})$. In terms of Nambu spinors 
\beq
\Psi = \left(\begin{array}{c}
\psi_{\uparrow}\\
\psi_{\downarrow}^\dagger
\end{array}\right), \quad \Psi^\dagger = \left(\psi_{\uparrow}^\dagger \,, \psi_\downarrow \right),
\eeq
that satisfy
\beq \label{antiNam}
\{\Psi_i(\mathbf{r}),\Psi_j^\dagger(\mathbf{r}') \} = \delta_{ij}\delta(\mathbf{r} - \mathbf{r}'),
\eeq
the Hamiltonian becomes
\beq
\begin{split}
H &= \int d^2r\, \Psi^\dagger\left(-\frac{\nabla^2}{2} + V(r) - \mu \right)\tau_3\Psi \\
 & + \int d^2r\, \Psi^\dagger \Delta(r)  \left( e^{i k \varphi/2}(p_x + i p_y)^\nu  e^{i k \varphi/2} \tau_+ + e^{-i k \varphi/2}(p_x - i p_y)^\nu  e^{-i k \varphi/2} \tau_- \right) \Psi .
 \end{split}
\eeq
The angular momentum and particle number operators are
\begin{align}
\hat{L}_z &= \int d^2r\, \Psi^\dagger \left(-i \frac{\partial}{\partial \varphi} \right) \Psi , \nonumber \\
\hat{N} &= \int d^2r\, \Psi^\dagger \tau_3 \Psi ,
\end{align}
respectively, where $\tau_i$ are the Pauli matrices and $\tau_{\pm} = \frac{1}{2}(\tau_1 \pm i \tau_2)$.
We work in polar coordinates, where 
\beq
(p_x \pm i p_y) = -i e^{\pm i \varphi} \left(\frac{\partial}{\partial r} \pm \frac{i}{r}\frac{\partial}{\partial \varphi} \right).
\eeq
In addition, the anti-commutation relations Eq.~\eqref{antiNam} lead to the following identity
\beq
[\Psi^\dagger \hat{A} \Psi,\Psi^\dagger \hat{B}\Psi] = \Psi^\dagger [\hat{A},\hat{B}] \Psi .
\eeq
Putting everything together, it is straightforward now to show that
\begin{align}
[\hat{L_z},\hat{H}] &= \Delta(r) \left[ e^{i k \varphi/2}(p_x + i p_y)^\nu  e^{i k \varphi/2} \tau_+ -  e^{-i k \varphi/2}(p_x - i p_y)^\nu  e^{- i k \varphi/2} \tau_-\right] \times \left(k + \nu \right) , \nonumber \\
[\hat{N},\hat{H}]   &= \Delta(r) \left[ e^{i k \varphi/2}(p_x + i p_y)^\nu  e^{i k \varphi/2} \tau_+ -  e^{-i k \varphi/2}(p_x - i p_y)^\nu  e^{- i k \varphi/2} \tau_-\right] \times 2 .
\end{align}
From this it follows that 
\beq
\label{genoam}
\mathscr{\hat{L}} = \hat{L}_z - \frac{k + \nu}{2} \hat{N}
\eeq
commutes with $\hat{H}$. The above calculation easily generalizes to the case of spin-triplet chiral SFs (where $\nu$ is odd) with MQVs, where the same operator Eq.~\eqref{genoam} is conserved.


\subsection{B. Construction of ground state wave function}

A generalized framework for deriving the ground state of a paired Hamiltonian through a Bogoliubov transformation was constructed in \cite{Labonte1974}. Here, we present a self-contained discussion and obtain the ground state wave functions for SFs with MQVs. While we focus on $s$-wave SFs here, this construction can be readily generalized to chiral SFs with vortices as well.

The eigenstates $(u,v)^T$ of the BdG Hamiltonian satisfy
\beq
\sum_{n'=1}^M H^{(l)}_{n,n'} \left(\begin{array}{c}
u_{n'm}^{(l)}\\
v_{n'm}^{(l)}
\end{array}
\right) = E_m^{(l)} \left(\begin{array}{c}
u_{nm}^{(l)}\\
v_{nm}^{(l)}
\end{array}
\right),
\eeq
where we have introduced a cut-off $M \gg 1$ on the radial quantum numbers $n,n'$. Suppose the number of positive and negative eigenvalues of $H^{(l)}$ are $M_+^{(l)}$ and $M_-^{(l)}$ respectively, with $M_+^{(l)} + M_-^{(l)} = 2 M$. In the absence of spectral asymmetry, $M_+^{(l)} = M_-^{(l)}$, but in general, $M_+^{(l)} \neq M_-^{(l)}$. Let us now order the energies such that $E_1^{(l)}\geq \dots \geq E_{2M}^{(l)}$, with
\beq
E_m^{(l)} > 0,\,\, m = 1,\dots, M_+^{(l)}, \quad \quad E_{m+M_+^{(l)}}^{(l)} < 0,\,\, m = 1,\dots, M_-^{(l)}.
\eeq

Next, we introduce the (inverse) Bogoliubov transformation
\begin{align}
\label{bgbtrans}
a_{n,l+k \uparrow} & = \sum_{m=1}^{M_+^{(l)}} u_{nm}^{(l)}\,b_m^{(l)} + \sum_{\bar{m}}^{M_-^{(l)}} \,u_{n,\bar{m} + M_+^{(l)}}^{(l)} d_{\bar{m}}^{(l)\dagger}, \nonumber \\
a_{n,-l \downarrow}^\dagger & = \sum_{m=1}^{M_+^{(l)}} v_{nm}^{(l)}\,b_m^{(l)} + \sum_{\bar{m}}^{M_-^{(l)}}\,v_{n,\bar{m} + M_+^{(l)}}^{(l)} d_{\bar{m}}^{(l)\dagger} .
\end{align}
Here, $b_m^{(l)}$ are Bogoliubov operators that annihilate a spin $\uparrow$ state with energy $E^{(l)}_m$ and $\mathscr{L}$-charge $l + k/2$. We can exploit the PH symmetry of the system to alternatively interpret $b_m^{(l)}$ as the creation operator for a spin $\downarrow$ state with energy $-E_m^{(l)}$ and $\mathscr{L}$-charge $-l - k/2$. In addition, we have introduced another set of Bogoliubov operators
\beq
d_{\bar{m}}^{(l)} \equiv b_{\bar{m} + M_+^{(l)}}^{(l)\dagger},\,\,\bar{m} = 1,\dots, M_-^{(l)},
\eeq
such that the operator $d_{\bar{m}}^{(l)}$ creates a spin $\uparrow$ state with energy $E^{(l)}_{\bar{m} + M_+^{(l)}} < 0$ and $\mathscr{L}$-charge $l + k/2$. As a matter of principle, we note that since the pairing Hamiltonian does not commute with the angular momentum operator $\hat{L}_z$, the Bogoliubov quasi-particles $b$ and $d$ carry $\mathscr{L}$-charge rather than an $l$ quantum number and thus the energy eigenvalues $E_k^{(l)}$ should be labelled instead by their $\mathscr{L}$ quantum number, $v = l - k/2$. However, since $v$ is simply $l$ shifted by a constant, it is more convenient to continue labelling the eigenvalues and quasi-particles by $l$. 

The BCS ground state is defined as the vacuum with respect to all positive energy quasi-particles, $\ket{BCS} \sim \otimes_l \ket{BCS}_l$, and must thus satisfy
\begin{align}
b_m^{(l)}\ket{BCS} &= 0\,\,(m = 1,\dots,M_+^{(l)}), \nonumber \\
d_{\bar{m}}^{(l)}\ket{BCS} &= 0\,\,(\bar{m} = 1,\dots,M_-^{(l)}).
\end{align}
When $M_+^{(l)} = M_-^{(l)}$, the ground state may be expressed as the state with all negative energy excitations occupied, with
\beq
\label{pair}
\ket{BCS}_l = \prod_{m=1}^{M} b_m^{(l)} \prod_{\bar{m}=1}^{M} d_{\bar{m}}^{(l)} \ket{0},
\eeq
where $\ket{0}$ is the Fock vacuum with respect to elementary fermions $a_{n,l\sigma}$. The paired nature of this state is evident in the wave function Eq.~\eqref{pair} as expressed in terms of the Bogoliubov operators, since a spin $\downarrow$ quasi-particle with $\mathscr{L}$-charge $-l - k/2$ (created by $b_m^{(l)}$) is paired with a spin $\uparrow$ quasi-particle with $\mathscr{L}$-charge $l + k/2$ (created by $d_{\bar{m}}^{(l)}$).

However, when $M_+^{(l)} \neq M_-^{(l)}$, there will be some states left unpaired as a consequence of the asymmetry in the spectrum. In order to elucidate the nature of the ground state in the presence of unpaired fermions, it is instructive to transform to a particular bases of elementary fermions and of Bogoliubov quasi-particles in which the structure of the ground state becomes especially transparent. Before presenting technical details of this procedure, we briefly describe the steps involved.
\begin{figure}[t]
\includegraphics[width=8cm]{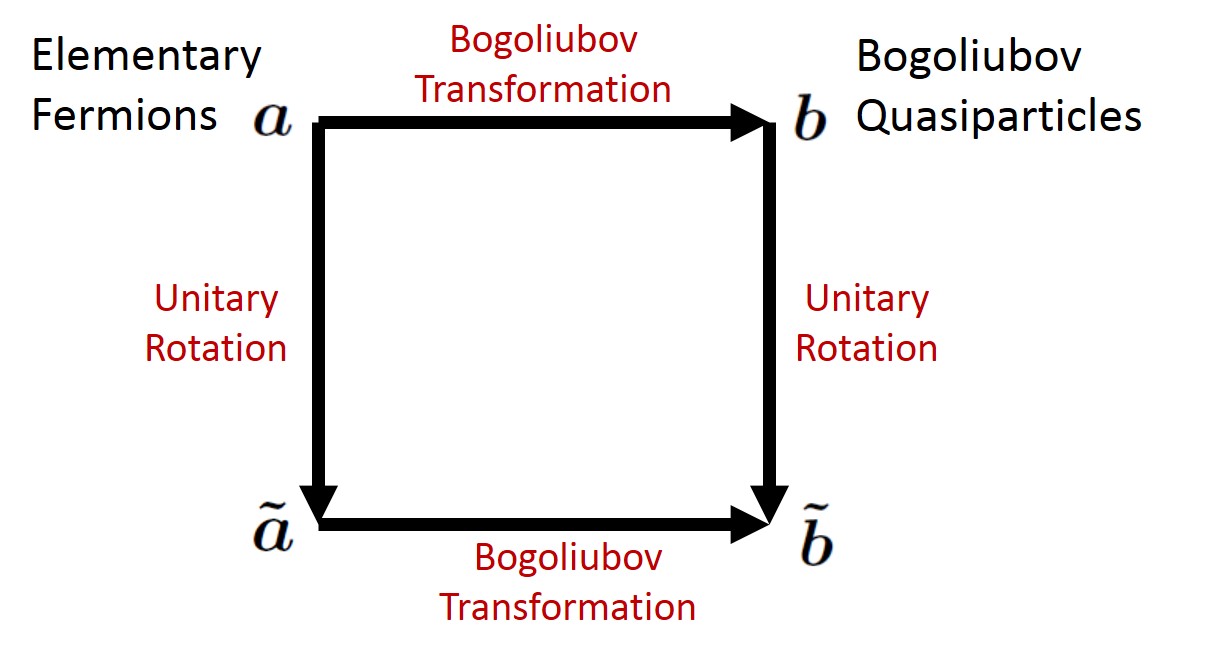}
\caption{Relations between different fermionic operators used in this section.}
\label{wavefunction}
\end{figure}

Fig.~\ref{wavefunction} illustrates the series of transformations that we perform in order to express $\ket{BCS}$ in a transparent form. First, we unitarily rotate the elementary fermions operators $a_{n,l\sigma}$ into a new basis of fermionic operators $\tilde{a}_{n,l\sigma}$ (where the $\tilde{a}$'s are linear combinations of only $a$'s but not $a^\dagger$'s). While the states destroyed by operators $\tilde{a}_{n,l\sigma}$ carry well defined spin and angular momentum quantum numbers, $\sigma$ and $l$ respectively, they are not energy eigenstates of the non-interacting Hamiltonian. In a similar spirit, we will unitarily rotate the quasi-particle operators $b,d$ into a new basis of quasi-particle operators $\tilde{b},\tilde{d}$. Importantly, since this operation does not mix quasi-particle creation and annihilation operators, the ground state is still defined as the vacuum with respect to positive energy excitations and hence satisfies $\tilde{b_m^{(l)}} \ket{BCS} = 0$ and $\tilde{d}_{\bar{m}}^{(l)} \ket{BCS} = 0$.  

Having established these new bases, we will then relate the rotated fermions $\tilde{a}$ to the rotated quasi-particles $\tilde{b},\tilde{d}$ through a Bogoliubov transformation akin to Eq.~\eqref{bgbtrans}. This relation then allows us to express $\ket{BCS}$ in terms of the rotated fermions $\tilde{a}$ in a manner that makes explicit the nature of pairing in the ground state since it naturally distinguishes between paired and unpaired states. We note that the purpose of these transformations is not to diagonalize the BdG Hamiltonian $H^{(l)}$ (which is diagonalized by the original Bogoliubov transformation Eq.~\eqref{bgbtrans}) but rather to explicitly construct the ground state wave function for an $s$-wave SF with an MQV.

We now discuss the above procedure in detail. To begin, we first invert Eq.~\eqref{bgbtrans} and write the unitary Bogoliubov transformation as
\beq
\left( \begin{array}{c}
b_{m}^{(l)}\\
d_{\bar{m}}^{(l)\dagger}
\end{array}
\right) = \sum_{n=1}^M \left(\begin{array}{cc}
S_{1,mn}^{(l)} & S_{2,mn}^{(l)}\\
S_{3,\bar{m}n}^{(l)} & S_{4,\bar{m}n}^{(l)}
\end{array}\right) \left( \begin{array}{c}
a_{n,l+k \uparrow}\\
a_{n,-l \downarrow}^\dagger
\end{array}
\right),
\eeq
where $S_1^{(l)}, S_2^{(l)}$ are $M_+^{(l)}\times M$ and $S_3^{(l)}, S_4^{(l)}$ are $M_-^{(l)}\times M$ dimensional matrices, respectively. Next, we perform a singular value decomposition on the matrices $S_2$ and $S_3$, $S_i^{(l)} = U_i^{(l)} \Sigma_i^{(l)} V_i^{(l)\dagger}$ where $U_i^{(l)}, V_i^{(l)}$ are unitary matrices and $\Sigma_i^{(l)}$ is a rectangular diagonal matrix with non-negative real entries. We then perform unitary rotations on the elementary fermions $a$ and the Bogoliubov quasi-particles, $b$ and $d$, 
\begin{align}
\tilde{a}_{j,l+k\uparrow} &= \sum_{n = 1}^{M} V_{3,jn}^{(l)\dagger}\,a_{n,l+k\uparrow} \quad j=1,\dots,M\, ,\nonumber \\
\tilde{a}_{j,-l\downarrow}^\dagger &= \sum_{n = 1}^{M} V_{2,jn}^{(l)\dagger}\,a_{n,-l\downarrow}^\dagger \quad j=1,\dots,M\, , \nonumber \\
\tilde{b}_{j}^{(l)} & = \sum_{n = 1}^{M_+^{(l)}} U_{2,jn}^{(l)\dagger}\,b_{n}^{(l)} \quad j=1,\dots,M_+^{(l)}\, , \nonumber \\
\tilde{d}_{j}^{(l)\dagger} & = \sum_{n = 1}^{M_-^{(l)}} U_{3,jn}^{(l)\dagger}\,d_{n}^{(l)\dagger} \quad j=1,\dots,M_-^{(l)}.
\end{align}
It is straightforward to check that these transformed operators satisfy the canonical anti-commutation relations. In this new basis, the Bogoliubov transformation is expressed as 
\beq
\label{newbog}
\left(\begin{array}{c}
\tilde{b}_m^{(l)} \\
\tilde{d}_{\bar{m}}^{(l)\dagger}
\end{array}
\right) = \sum_{m=1}^M \left( \begin{array}{cc}
\Lambda_{1,mn}^{(l)} & \Sigma_{2,mn}^{(l)} \\
\Sigma_{3,\bar{m}n}^{(l)} & \Lambda_{4,\bar{m}n}^{(l)}
\end{array}
\right) \left(\begin{array}{c}
\tilde{a}_{n,l+k \uparrow}\\
\tilde{a}_{n,-l \downarrow}^\dagger
\end{array}
\right),
\eeq
where $\Lambda_{1}^{(l)} = U_2^{(l)\dagger} S_1^{(l)} V_3^{(l)}$ and $\Lambda_{4}^{(l)} = U_3^{(l)\dagger} S_4^{(l)} V_2^{(l)}$. Before proceeding with the construction of the ground state, it is necessary to establish some properties of $\Sigma_2^{(l)}$ and $\Sigma_3^{(l)}$.
In particular, we will show now that in the presence of a non-trivial spectral asymmetry $\eta_l = M_+^{(l)} - M_-^{(l)}$, either one of $\Sigma_2^{(l)}$ or $\Sigma_3^{(l)}$ have $|\eta_l|/2$ entries on the diagonal which are equal to one. To prove this, let us assume without loss of generality that $M_+^{(l)} < M < M_-^{(l)}$. Since the Bogoliubov transformation is unitary, the transformation matrix $S^{(l)} \equiv \left(\begin{array}{cc}S^{(l)}_1 & S^{(l)}_2 \\ S^{(l)}_3 & S^{(l)}_4 \end{array} \right)$ satisfies $S^{(l)\dagger} S^{(l)} = S^{(l)} S^{(l)\dagger} = \mathbb{I}_{2M \times 2M}$. This leads to the conditions
\begin{align}
\label{unitary}
S_1^{(l)} S_1^{(l)\dagger} + S_2^{(l)} S_2^{(l)\dagger} &= \mathbb{I}_{M_+^{(l)} \times M_+^{(l)}}\, ,\nonumber \\
S_1^{(l)\dagger} S_1^{(l)} + S_3^{(l)\dagger} S_3^{(l)} &= \mathbb{I}_{M\times M}.
\end{align}
Since $S_3^{(l)\dagger} S_3^{(l)}$ and $S_2^{(l)} S_2^{(l)\dagger}$ are Hermitian matrices, they have real eigenvalues and eigenvectors,
\begin{align}
\left(S_2^{(l)} S_2^{(l)\dagger} \right) y_{2j}^{(l)} &= \lambda_{2j}^{(l)} y_{2j}^{(l)} \quad j=1,\dots,M_+^{(l)}\, , \nonumber \\
\left(S_3^{(l)\dagger} S_3^{(l)} \right) y_{3j}^{(l)} &= \lambda_{3j}^{(l)} y_{3j}^{(l)} \quad j=1,\dots,M .
\end{align}
Eq.~\eqref{unitary} then implies 
\begin{align}
\left(S_1^{(l)} S_1^{(l)\dagger} \right) y_{2j}^{(l)} &= \left(1 -\lambda_{2j}^{(l)}\right) y_{2j}^{(l)} \quad j=1,\dots,M_+^{(l)}\,, \nonumber \\
\left(S_1^{(l)\dagger} S_1^{(l)} \right) y_{3j}^{(l)} &= \left(1 -\lambda_{3j}^{(l)}\right) y_{3j}^{(l)} \quad j=1,\dots,M .
\end{align}
The singular values of $S_1^{(l)}$, however, are the square roots of the non-zero eigenvalues of both $S_1^{(l)\dagger} S_1^{(l)}$ and $S_1^{(l)} S_1^{(l)\dagger}$. Since $S_1^{(l)\dagger} S_1^{(l)}$ has exactly $M - M_+^{(l)}$ more eigenvalues than $S_1^{(l)} S_1^{(l)\dagger}$, those extra eigenvalues must necessarily be zero. This in turn implies that $S_3^{(l)\dagger} S_3^{(l)}$ has $M_\uparrow^{(l)} = M - M_+^{(l)}$ unity eigenvalues, or equivalently, that $\Sigma^{(l)}_3$ has precisely $M_\uparrow^{(l)}$ unity entries on the diagonal. In addition, since $\Sigma^{(l)}_3$ is a rectangular diagonal $M_-^{(l)}\times M$ matrix, its last $M^{(l)}_- - M = M - M_+^{(l)}= M_\uparrow^{(l)}$ rows contain only zeros.

Following this discussion, in general for any $M_+^{(l)}$ and $M_-^{(l)}$ we can define the quantities $M_\uparrow^{(l)} = \text{max}\left(M - M_+^{(l)} ,0\right)$ and $M_\downarrow^{(l)} = \text{max}\left(M - M_-^{(l)} ,0\right)$. Using the fact that $M_+^{(l)} + M_-^{(l)} = 2M$, we can also show that $M_+^{(l)} = M + M_\downarrow^{(l)} - M_\uparrow^{(l)}$ and $M_-^{(l)} = M - M_\downarrow^{(l)} + M_\uparrow^{(l)}$. From this, it follows that
\beq
\eta_l = M_+^{(l)} - M_-^{(l)} = 2(M_\downarrow^{(l)} - M_\uparrow^{(l)}) = \begin{cases}
2(M - M_-^{(l)})>0, &\quad M_+^{(l)}>M>M_-^{(l)} \\
0, &\quad M_-^{(l)}=M=M_+^{(l)} \\
2(M_+^{(l)} - M)<0, &\quad M_+^{(l)}<M<M_-^{(l)}
\end{cases}.
\eeq

Combining the above results with the unitarity of the Bogoliubov transformation Eq.~\eqref{newbog}, we conclude that in the presence of spectral asymmetry we will have cases where $\tilde{b}^{(l)} = \tilde{a}_{-l\downarrow}^\dagger$ or $\tilde{d}^{(l)} = \tilde{a}_{l+k \uparrow}^\dagger$, which will give rise to unpaired fermions in the ground state. 

Specifically, we find that the Bogoliubov transformation Eq.~\eqref{newbog} splits into three classes
\begin{align} \label{Bt1}
\tilde{b}_m &= \tilde{a}_{m,-l\downarrow}^{\dagger} \quad m = 1,\dots,M_\downarrow \, ,\nonumber \\
\tilde{d}_{\bar{m}}^\dagger &= \tilde{a}_{\bar{m},l+k\uparrow} \quad \bar{m} = 1,\dots,M_\uparrow\, ,
\end{align}
\begin{align} \label{Bt2}
\tilde{b}_{M_\downarrow + m} &= \sum_{n=1}^{M - M_\uparrow} \left(\Lambda_1\right)_{M_\downarrow + m, M_\uparrow + n} \tilde{a}_{M_\uparrow + n, l+k \uparrow} + \sum_{n=1}^{M - M_\downarrow} \left(\Sigma_2\right)_{M_\downarrow + m, M_\downarrow + n} \tilde{a}_{M_\downarrow + n, -l \downarrow}^\dagger, \quad m = 1,\dots,M - M_\downarrow- M_\uparrow \, ,\nonumber \\
\tilde{d}_{M_\uparrow + m}^\dagger &= \sum_{n=1}^{M - M_\uparrow} \left(\Sigma_3\right)_{M_\uparrow + m, M_\uparrow + n} \tilde{a}_{M_\uparrow + n, l+k \uparrow} + \sum_{n=1}^{M - M_\downarrow} \left(\Lambda_4\right)_{M_\uparrow + m, M_\downarrow + n} \tilde{a}_{M_\downarrow + n, -l \downarrow}^\dagger, \quad m = 1,\dots,M - M_\downarrow- M_\uparrow \, ,
\end{align}
and
\begin{align} \label{Bt3}
\tilde{b}_{M-M_\uparrow + m} &= \sum_{n=1}^{M - M_\uparrow} \left(\Lambda_1\right)_{M-M_\uparrow + m, M_\uparrow + n} \tilde{a}_{M_\uparrow + n, l+k \uparrow}, \quad m = 1,\dots,M_\downarrow \, ,\nonumber \\
\tilde{d}_{M-M_\downarrow + m}^\dagger &=  \sum_{n=1}^{M - M_\downarrow} \left(\Lambda_4\right)_{M-M_\downarrow + m, M_\downarrow + n} \tilde{a}_{M_\downarrow + n, -l \downarrow}^\dagger, \quad m = 1,\dots, M_\uparrow \, ,
\end{align}
where we have omitted the $(l)$ superscript to simplify notation.
Physically, these three classes correspond to three different kinds of quasi-particle operators \cite{Ring}:
\begin{itemize}
\item Occupied levels Eq.~\eqref{Bt1} are those where $\tilde{b} = \tilde{a}_\downarrow^\dagger$ or $\tilde{d} = \tilde{a}_\uparrow^\dagger$. These operators create a unitarily rotated fermion with unit probability. 
\item Paired levels Eq.~\eqref{Bt2} are those for which $\tilde{b}$ and $\tilde{d}$ are non-trivial superpositions of $\tilde{a}_\uparrow$ and $\tilde{a}_\downarrow^\dagger$. 
\item Empty levels Eq.~\eqref{Bt3} $\tilde{b}$ and $\tilde{d}$ are linear superpositions of $\tilde{a}_\uparrow$'s and $\tilde{a}_\downarrow$'s, respectively. These operators annihilate the Fock vacuum $\ket{0}$. 
\end{itemize}

The ground state is still the vacuum with respect to all positive energy quasi-particles and, in terms of the unitarily rotated Bogoliubov quasi-particles, it is given by $\ket{BCS} \sim \otimes_l \ket{BCS}_l$ with
\beq \label{GSnew}
\ket{BCS}_l = \sideset{}{'}\prod_{m} \tilde{b}^{(l)}_m \sideset{}{'}\prod_{\bar{m}} \tilde{d}^{(l)}_{\bar{m}} \ket{0}
\eeq
where the restricted product runs over all paired and occupied levels. Since empty levels annihilate the Fock vacuum, they are not included in Eq.~\eqref{GSnew}. By construction this state satisfies
\beq
\tilde{b}^{(l)}_m \ket{BCS}_l = 0, \quad \tilde{d}^{(l)}_{\bar{m}} \ket{BCS}_l = 0 ,
\eeq
for all $m=1,\dots, M_+^{(l)}$ and $\bar m=1,\dots, M_-^{(l)}$.
We can further simplify the ground state since it factorizes into unpaired and paired terms,
\beq
\ket{BCS}_l = \underbrace{\left(\prod_{j=1}^{M_\downarrow^{(l)}} \tilde{a}_{j,-l\downarrow}^\dagger\right) \left(\prod_{j=1}^{M_\uparrow^{(l)}} \tilde{a}_{j,l+k\uparrow}^\dagger \right)}_{\text{unpaired}} \underbrace{\left(\sideset{}{'}\prod_{m>M_\uparrow^{(l)}}^M \sideset{}{'}\prod_{\bar{m}>M_\downarrow^{(l)}}^M \tilde{b}_m^{(l)} \tilde{d}_{\bar{m}}^{(l)} \right)}_{\text{paired}} \ket{0}.
\eeq
The first two terms in the product clearly indicate that for $M_\sigma^{(l)}\neq 0 $, there are unpaired fermions in the ground state. Following \cite{Labonte1974}, we can now write the paired part of the ground state in terms of creation operators $\tilde a^{\dagger}$ of elementary (unitarily rotated) fermions
\beq
\label{wfnsm}
\ket{BCS}_l = \left(\prod_{j=1}^{M_\downarrow^{(l)}} \tilde{a}_{j,-l\downarrow}^\dagger\right) \left(\prod_{j=1}^{M_\uparrow^{(l)}} \tilde{a}_{j,l+k\uparrow}^\dagger \right) \exp\left(\sum_{j>M_\uparrow^{(l)}}^M \sum_{j'>M_\downarrow^{(l)}}^M \tilde{a}_{j,l+k \uparrow}^\dagger \mathcal{K}^{(l)}_{j,j'} \tilde{a}_{j',-l\downarrow}^\dagger \right) \ket{0} ,
\eeq 
where the kernel $\mathcal{K}$ satisfies 
\beq
\sum_{j'>M_\uparrow^{(l)}}^M \Lambda_{1,jj'}^{(l)} \mathcal{K}_{j',j''}^{(l)} = -\Sigma_{2,jj''}^{(l)}\,,
\eeq
with $j,j'' > M_\downarrow^{(l)}$. 

In order to derive Eq.~\eqref{conserved} in the main text, we note that we can express the conserved operator $\mathscr{\hat{L}} = \hat{L} - k \hat{N}/2$ in terms of the elementary fermions as
\beq
\mathscr{\hat{L}} = \sum_{nl\sigma} \left(l - \frac{k}{2} \right) a_{n,l\sigma}^\dagger a_{n,l\sigma} = \sum_{nl\sigma} \left(l - \frac{k}{2} \right) \tilde{a}_{n,l\sigma}^\dagger \tilde{a}_{n,l\sigma},
\eeq
where the last equality follows since the $\tilde{a}$'s are unitarily related to the $a$'s. Since the set of operators that appear in the exponential part of $\ket{BCS}_l$ anti-commute with those that appear in the products (unpaired fermions), we can consider the action of $\mathscr{\hat{L}}$ on these separately. First, we evaluate the commutator
\beq
\left[ \mathscr{\hat{L}}, \prod_{j=1}^{M_\downarrow^{(l)}} \tilde{a}_{j,-l\downarrow}^\dagger\right] = 
- M_\downarrow^{(l)} \left( l + \frac{k}{2}\right)\left(\prod_{j=1}^{M_\downarrow^{(l)}} \tilde{a}_{j,-l\downarrow}^\dagger\right),
\eeq
and similarly, 
\beq
\left[ \mathscr{\hat{L}}, \prod_{j=1}^{M_\uparrow^{(l)}} \tilde{a}_{j,l+k\uparrow}^\dagger \right] = M_\uparrow^{(l)} \left(l + \frac{k}{2}\right) \left(\prod_{j=1}^{M_\uparrow^{(l)}} \tilde{a}_{j,l+k\uparrow}^\dagger \right),
\eeq
which follows from the usual anti-commutation relations satisfied by the rotated fermions $\tilde{a}$. Next, we consider the action of $\mathscr{\hat{L}}$ on the exponential (paired) sector of the wave function. This is done by first calculating the contribution from the spin $\uparrow$ sector, 
\beq
\left[\sum_{nl'} \left(l' - \frac{k}{2} \right) \tilde{a}_{n,l'\uparrow}^\dagger \tilde{a}_{n,l'\uparrow}\, , \left(\sum_{j>M_\uparrow^{(l)}}^M \sum_{j'>M_\downarrow^{(l)}}^M \tilde{a}_{j,l+k \uparrow}^\dagger \mathcal{K}^{(l)}_{j,j'} \tilde{a}_{j',-l\downarrow}^\dagger \right) \right] 
= \left(l + \frac{k}{2} \right) \sum_{j>M_\uparrow^{(l)}}^M \sum_{j'>M_\downarrow^{(l)}}^M \tilde{a}_{j,l+k \uparrow}^\dagger \mathcal{K}^{(l)}_{j,j'} \tilde{a}_{j',-l\downarrow}^\dagger ,
\eeq
and then observing that the contribution from the spin $\downarrow$ sector
\beq
\left[\sum_{nl'} \left(l' - \frac{k}{2} \right) \tilde{a}_{n,l'\downarrow}^\dagger \tilde{a}_{n,l'\downarrow}, \left(\sum_{j>M_\uparrow^{(l)}}^M \sum_{j'>M_\downarrow^{(l)}}^M \tilde{a}_{j,l+k \uparrow}^\dagger \mathcal{K}^{(l)}_{j,j'} \tilde{a}_{j',-l\downarrow}^\dagger \right) \right] = - \left(l + \frac{k}{2} \right) \sum_{j>M_\uparrow^{(l)}}^M \sum_{j'>M_\downarrow^{(l)}}^M \tilde{a}_{j,l+k \uparrow}^\dagger \mathcal{K}^{(l)}_{j,j'} \tilde{a}_{j',-l\downarrow}^\dagger,
\eeq
exactly compensates for that coming from the spin $\uparrow$ sector. Hence, we see that $\mathscr{\hat{L}}$ commutes with the exponential part of the wave function and so the paired levels do not contribute to $\mathscr{L}$,
\beq
\left[ \mathscr{\hat{L}}, \exp \left(\sum_{j>M_\uparrow^{(l)}}^M \sum_{j'>M_\downarrow^{(l)}}^M \tilde{a}_{j,l+k \uparrow}^\dagger \mathcal{K}^{(l)}_{j,j'} \tilde{a}_{j',-l\downarrow}^\dagger \right) \right] = 0.
\eeq
Since the $\tilde{a}$'s are unitarily related to the $a$'s, $\ket{0}$ is also the Fock vacuum with respect to $\tilde{a}_{j,l\sigma}$. Putting the above together, we hence find that the eigenvalue $\mathscr{L}$ of the operator $\mathscr{\hat{L}}$ when evaluated in the ground state $\ket{BCS} \sim \otimes_l \ket{BCS}_l$, with $\ket{BCS}_l$ given by Eq.~\eqref{wfnsm}, is
\beq
\mathscr{L} = \sum_l \left(l + \frac{k}{2} \right) \left(M_\uparrow^{(l)} - M_\downarrow^{(l)}\right) = -\frac{1}{2} \sum_l \left(l + \frac{k}{2} \right)\eta_l ,
\eeq
where $\eta_l = 2(M_\downarrow^{(l)} - M_\uparrow^{(l)})$.


\subsection{C. Analytic solution for vortex core states}

We generalize the CdGM method~\cite{Caroli1964} for obtaining the vortex core bound states in an $s$-wave superconductor to the case of multiply quantized vortices. For a step-like pair-potential, an explicit solution was obtained previously in~\cite{Bardeen1969,Berthod2005}. The procedure presented here applies more generally to any pairing term $\Delta(r)$ and agrees with that calculation where the regimes of validity overlap. 

We start with an $s$-wave state with a vortex of vorticity $k$. The (symmetric) gap function is thus $\Delta(r) e^{i k \varphi}$ where $\lim_{r\to0}\Delta(r)=0$ and $\lim_{r\to \infty}\Delta(r)=\Delta_0$. The BdG equations are hence
\beq
\label{BdG}
\left(
\begin{array}{cc}
-\frac{1}{2}\nabla^2 - \mu & \Delta(r) e^{i k \varphi} \\
\Delta(r) e^{-i k \varphi} & \frac{1}{2}\nabla^2 + \mu
\end{array}
\right)
\left(
\begin{array}{c}
u \\
v
\end{array}
\right) = E \left(
\begin{array}{c}
u \\
v
\end{array} \right) .
\eeq
Separating the angular and radial dependence of the BdG solutions, we let 
\begin{align}
u &= u(r) e^{i \left(l + \frac{k}{2} \right) \varphi} ,\nonumber \\
v &= v(r) e^{i \left(l - \frac{k}{2} \right) \varphi} ,
\end{align}
where $l \in \mathbb{Z} \left(\mathbb{Z} + \frac{1}{2}\right)$ if $k \in \text{Even(Odd)}$, such that Eq.~\eqref{BdG} becomes
\begin{align}
-\frac{1}{2}\left(\frac{\partial^2}{\partial r^2} + \frac{1}{r} \frac{\partial}{\partial r} - \frac{\left(l + \frac{k}{2}\right)^2}{r^2} + 2 \mu \right)u(r) +\Delta(r) v(r) &= E u(r)\,,\nonumber \\
\frac{1}{2}\left(\frac{\partial^2}{\partial r^2} + \frac{1}{r} \frac{\partial}{\partial r} - \frac{\left(l - \frac{k}{2}\right)^2}{r^2} + 2 \mu \right)v(r) +\Delta(r) u(r) &= E v(r)\,.
\end{align}
We now rewrite these equations in the form
\begin{align}
-\frac{1}{2}\left(\frac{\partial^2}{\partial r^2} + \frac{1}{r} \frac{\partial}{\partial r} - \frac{\alpha^2}{r^2} + 2 \mu \right)u(r) +\Delta(r) v(r) &= \left(E - \frac{\beta}{2 r^2} \right) u(r)\,,\nonumber \\
\frac{1}{2}\left(\frac{\partial^2}{\partial r^2} + \frac{1}{r} \frac{\partial}{\partial r} - \frac{\alpha^2}{r^2} + 2 \mu \right)v(r) +\Delta(r) u(r) &= \left(E - \frac{\beta}{2 r^2} \right) v(r)\,,
\end{align}
where we have defined $\alpha = \sqrt{l^2 + \frac{k^2}{4}}$ and $\beta = lk$.

In order to derive an analytic solution for these coupled equations, we introduce a radius $r = r^*$ such that $\frac{1}{k_F}\ll r^*\ll \xi$, where $k_F= \sqrt{2 \mu}$ is the Fermi momentum and $\xi = k_F/\Delta_0$ is the coherence length. We then consider the BdG equations~\eqref{BdG} separately in the limits where $r \ll r^*$ and where $r \gg r^*$. Demanding that the wave function be continuous, we then match the solutions from these two regimes at $r = r^*$ to arrive at a solution that holds over the entire range.   

We first consider the limit where $r \ll r^*$. For physically relevant pairing terms, $\Delta(r)\to 0$ in this limit and we are hence justified in ignoring the pairing term. The BdG equations~\eqref{BdG} thus decouple in this limit,
\begin{align}
\left(\frac{\partial^2}{\partial r^2} + \frac{1}{r} \frac{\partial}{\partial r} - \frac{\left(l + \frac{k}{2}\right)^2}{r^2} + 2 (\mu + E)\right)u &= 0 \,,\nonumber \\
\left(\frac{\partial^2}{\partial r^2} + \frac{1}{r} \frac{\partial}{\partial r} - \frac{\left(l - \frac{k}{2}\right)^2}{r^2} + 2 (\mu - E)\right)v &= 0  \,.
\end{align}
The solutions to these equations are Bessel functions parametrized by $\sqrt{2(\mu \pm E)}$. Since we are interested in understanding the nature of the vortex core states close to zero energy, we make an additional approximation and consider energies such that $E \ll \mu$. Reinstating all the proper units, this corresponds to
\beq
\left(\mu \pm E\right)^{1/2} = \sqrt{\frac{\hbar^2}{2 m_e}} \left(k_F^2 \pm \frac{2 m_e E}{\hbar^2}\right)^\frac{1}{2} \approx \sqrt{\frac{\hbar^2}{2 m_e}} \left(k_F \pm p\right)\,,
\eeq
where we have defined $p = \frac{E m_e}{\hbar^2 k_F}$. Thus, the solutions for $r \ll r^*$ are
\begin{align}
\label{smallr}
u(r) &= C_1 J_{l+ \frac{k}{2}}\left((k_F+p)r\right) \,,\nonumber \\
v(r) &= C_2 J_{l- \frac{k}{2}}\left((k_F-p)r\right) \,,
\end{align}
where $C_1$ and $C_2$ are arbitrary constants and where the Bessel functions of the second kind $Y$ are chosen to have vanishing amplitudes since we require a well behaved solution in the limit $r\to 0$.

Next, we consider the case where $r \gg r^*$. In this limit, we expect that the pairing term is approximately constant $\Delta(r) \to \Delta_0$. Hence, we write the BdG solutions as rapidly oscillating Hankel functions enveloped by functions that vary slowly i.e.,
\beq
\left(
\begin{array}{c}
u(r) \\
v(r)
\end{array}\right) = 
\left(
\begin{array}{c}
f(r) H^{(1)}_{\alpha}(k_F r) \\
g(r) H^{(1)}_{\alpha}(k_F r)
\end{array} 
\right)
+
\left(
\begin{array}{c}
\tilde{f}(r) H^{(2)}_{\alpha}(k_F r) \\
\tilde{g}(r) H^{(2)}_{\alpha}(k_F r)
\end{array} 
\right)\,,
\eeq
where $H^{(1,2)}(r)$ are Hankel functions of the first and second kind. Substituting this ansatz into the BdG equations~\eqref{BdG} and considering only the $H^{(1)}$ component of the solution (since the other follows from this immediately), we find  
\begin{align}
-\frac{1}{2}\left(f'' H_{\alpha} +2 f' H'_{\alpha} + \frac{1}{r} H_{\alpha} f' \right) + \Delta H_{\alpha} g &= \left(E - \frac{\beta}{2 r^2} \right) f H_{\alpha} \,,\nonumber \\
\frac{1}{2}\left(g'' H_{\alpha} +2 g' H'_{\alpha} + \frac{1}{r} H_{\alpha} g' \right) + \Delta H_{\alpha} f &= \left(E - \frac{\beta}{2 r^2} \right) g H_{\alpha} \,.
\end{align}
Here, $H_\alpha \equiv H^{(1)}_\alpha$. Since we are in the regime $k_F r \gg 1$, we use the asymptotic expansion for the Hankel functions
\beq
\frac{\partial}{\partial r}H^{(1)}_{\alpha}(k_F r) \approx i k_F H^{(1)}_{\alpha}(k_F r) \,,
\eeq
and drop the terms $f'',g'',\frac{f'}{r},\frac{g'}{r}$, which is justified since we are considering the limit $r \gg \frac{1}{k_F}$. With these further approximations, we find that the slowly varying envelope functions $f,g$ satisfy
\begin{align}
\label{sloweq}
-i k_F f' + \Delta g &= \left(E - \frac{\beta}{2 r^2} \right)f \,,\nonumber \\
i k_F g' + \Delta f &= \left(E - \frac{\beta}{2 r^2} \right)g \,.
\end{align}
To solve these coupled equations, we treat the right hand side as a perturbation. To zeroth order, 
\begin{align}
-i k_F f' + \Delta g &= 0 \,,\nonumber \\
i k_F g' + \Delta f &= 0 \,.
\end{align}
Imposing the condition that the solutions remain well behaved as $r \to \infty$, we find that the solutions are 
\beq
\left(
\begin{array}{c}
f \\
g
\end{array}
\right) = B\,\,\text{exp}\left(-\frac{1}{k_F} \int_{0}^{r} dr' \, \Delta(r') \right) 
\left(
\begin{array}{c}
1 \\
-i
\end{array}
\right) \,.
\eeq
In order to find the solution to first order in $E$, we make the ansatz
\beq
\left(
\begin{array}{c}
f \\
g
\end{array}
\right) = 
B\,\,\text{exp}\left(-\frac{1}{k_F} \int_{0}^{r} dr' \, \Delta(r') \right) \left(
\begin{array}{c}
e^{i\psi(r)} \\
-i e^{-i\psi(r)}
\end{array} 
\right) \, ,
\eeq
which when substituted into Eq.~\eqref{sloweq} leads to
\beq
k_F \psi' - 2 \Delta \sin(\psi) = \left(E - \frac{\beta}{2 r^2} \right) \,. 
\eeq
Approximating $\sin(\psi) \approx \psi$, this is equivalent to 
\beq
k_F \psi' - 2\Delta \psi = \left(E - \frac{\beta}{2 r^2} \right) \,,
\eeq
the solution to which is 
\beq
\label{psi}
\psi(r) = -\frac{e^{\frac{2}{k_F}\int_0^r dr' \, \Delta(r')}}{k_F}\int_r^\infty dr'\, \left(E - \frac{\beta}{2 r'^2} \right) e^{ - \frac{2}{k_F}\int_0^{r'} dr''\, \Delta(r'')} \,.
\eeq
With this, we find that the leading order solution of Eq.~\eqref{BdG} in the limit $r \gg r^*$ is 
\beq
\label{larger}
\left(
\begin{array}{c}
u(r) \\
v(r)
\end{array}
\right) = 
\left(
\begin{array}{c}
B_1 e^{i \psi(r)} H_\alpha^{(1)}(k_F r) + B_2 e^{-i \psi(r)} H_\alpha^{(2)}(k_F r) \\
-i B_1 e^{-i \psi(r)} H_\alpha^{(1)}(k_F r) +i B_2 e^{i \psi(r)} H_\alpha^{(2)}(k_F r)
\end{array}
\right)\times \exp\left(-\frac{1}{k_F}\int_0^r dr'\,\Delta(r')\right),
\eeq
with $\psi(r)$ given by Eq.~\eqref{psi}. 

We have thus constructed the general solution of the BdG equations~\eqref{BdG}, with the solution in the limits $r \ll r^*$ and $r \gg r^*$ given by Eq.~\eqref{smallr} and Eq.~\eqref{larger} respectively. In order to completely fix the undetermined coefficients and to get the quantization condition on the energy, we demand that the solution be continuous and match the solution in the regime $r<r^*$ with that in the regime $r>r^*$. 

Since we wish to extend the solution for $r\ll r^*$ Eq.~\eqref{smallr} to the vicinity of $r^*$, we use the asymptotic expansion
\beq
J_\nu(k_F r) \approx \sqrt{\frac{2}{\pi k_F r}} \cos\left(k_F r + \frac{\nu^2 -\frac{1}{4}}{2 k_F r} - \frac{2\nu+1}{4}\pi \right) \,,
\eeq
such that Eq.~\eqref{smallr} becomes
\begin{align}
\label{asymp1}
u(r) &\approx C_1 \sqrt{\frac{2}{\pi k_F r}} \cos\left((k_F + p)r + \frac{\left(l + \frac{k}{2}\right)^2 -\frac{1}{4}}{2 k_F r} - \frac{2\left(l + \frac{k}{2} \right)+1}{4}\pi \right) \,,\nonumber \\
v(r) &\approx C_2 \sqrt{\frac{2}{\pi k_F r}} \cos\left((k_F - p)r + \frac{\left(l - \frac{k}{2}\right)^2 -\frac{1}{4}}{2 k_F r} - \frac{2\left(l - \frac{k}{2} \right)+1}{4}\pi \right) \,.
\end{align}

Similarly, we want to extend the solution from the opposite regime ($r \gg r^*$) towards the vicinity of $r^*$ and we use the asymptotic expansion
\beq
H^{(1),(2)}_\nu(k_F r) \approx \sqrt{\frac{2}{\pi k_F r}} \text{exp}\left[i\left(k_F r \pm \frac{\nu^2 - \frac{1}{4}}{2 k_F r} \mp \frac{2\nu+1}{4}\pi \right)\right] \,,
\eeq
such that Eq.~\eqref{larger} becomes 
\beq
\label{asymp2}
\left(
\begin{array}{c}
u(r) \\
v(r)
\end{array}
\right) \approx \sqrt{\frac{2}{\pi k_F r}} e^{-\frac{1}{k_F}\int_0^r dr'\,\Delta(r')}
\left(
\begin{array}{c}
B_1 e^{i \psi(r) + i \gamma(r)} + B_2 e^{-i \psi(r) - i \gamma(r)} \\
-i B_1 e^{-i \psi(r) + i \gamma(r)} +i B_2 e^{i \psi(r) - i \gamma(r)}
\end{array}
\right) \,,
\eeq
where $\gamma(r) = k_F r + \frac{\alpha^2 - \frac{1}{4}}{2 k_F r} - \frac{2 \alpha + 1}{4}\pi$. We must now match the solutions from Eqs.~\eqref{asymp1} and \eqref{asymp2} at $r \sim r^*$ in order to find a continuous solution.

We first match $u(r)$. Making the ansatz $B_1 = \frac{C_1}{2}e^{i \kappa}$ and $B_2 = \frac{C_1}{2}e^{-i \kappa}$ and matching the solutions at $r \sim r^*$ leads to a condition on $\psi(r)$
\beq
\label{match1}
\psi(r^*) + \kappa - \frac{E r^*}{k_F} + \left(l - \alpha + \frac{k}{2}\right) \frac{\pi}{2} - \frac{\beta}{2 k_F r^*} = 0 \,.
\eeq
Next, we match $v(r)$, which leads to
\beq
\label{match2}
\psi(r^*) - \kappa - \frac{E r^*}{k_F} - \left(l - \alpha - \frac{k}{2} \right) \frac{\pi}{2} - \frac{\beta}{2 k_F r^*} + \left(n + \frac{1}{2} \right)\pi = 0 \,.
\eeq
Here, $n \in \mathbb{Z}$. Comparing Eqs.~\eqref{match1} and \eqref{match2}, we find that the parameter $\kappa$ is
\beq
\kappa = (\alpha-l)\frac{\pi}{2} + \left(n + \frac{1}{2}\right)\frac{\pi}{2} \,.
\eeq
Thus, we find that at $r \sim r^*$ the function $\psi(r)$ is approximately
\beq
\label{psiapprox}
\psi(r^*) \sim \frac{E r^*}{k_F} + \frac{\beta}{2 k_F r^*} - \frac{k + 1}{4}\pi - \frac{n}{2}\pi .
\eeq
However, recall that we earlier found that $\psi(r)$ is given by Eq.~\eqref{psi} while constructing the BdG solution in the regime $r \gg r^*$. Hence, in order to have a consistent solution, we must compare the approximate solution Eq.~\eqref{psiapprox} with Eq.~\eqref{psi}.
Since we want to understand the behavior of $\psi(r)$ in the regime where $r \sim r^* \gg \frac{1}{k_F}$, as a first approximation we can drop the term $e^{\frac{2}{k_F}\int_0^r dr' \, \Delta(r')}$. This is justified since the pairing term $\Delta(r)$ approaches 0 at least linearly and, as we are working in the regime where $r\ll \xi$, we can approximate this term as $e^{\frac{\Delta_0 r}{k_F}} = e^{\frac{r}{\xi}} \sim 1$. Thus, in order to understand the behavior of $\psi(r)$ in the vicinity of $r^*$ we need only consider the integral
\beq
\label{approx}
I = \int_r^\infty dr' \, e^{-\frac{2}{k_F}\int_0^{r'} dr'' \Delta(r'')} \left(E - \frac{\beta}{2 r'^2} \right) \,.
\eeq
We write $I = I_1 + I_2$ where
\begin{align}
I_1 &= E \int_r^\infty dr' \, e^{-\frac{2}{k_F}\int_0^{r'} dr'' \Delta(r'')} ,\nonumber \\
I_2 &= - \frac{\beta}{2}\int_r^\infty \frac{dr'}{r'^2} \, e^{-\frac{2}{k_F}\int_0^{r'} dr'' \Delta(r'')} .
\end{align}
We can now approximate $I_1$ as
\beq
I_1 \sim -E r + E\int_0^\infty dr' \, e^{-\frac{2}{k_F}\int_0^{r'} dr'' \Delta(r'')}   ,
\eeq
and can further write
\beq
I_2 = \frac{\beta}{2} \int_r^\infty dr' \, \frac{\partial}{\partial r'}\left(\frac{1}{r'}\right)e^{-\frac{2}{k_F}\int_0^{r'} dr'' \Delta(r'')} .
\eeq
Integrating $I_2$ by parts, we find
\beq
I_2 = \frac{\beta}{2} \left[\frac{e^{-\frac{2}{k_F}\int_0^{r'} dr'' \, \Delta(r'')}}{r'} \right]_r^\infty + \frac{\beta}{k_F} \int_r^\infty \frac{dr'}{r'} \Delta(r') e^{-\frac{2}{k_F}\int_0^{r'} dr'' \, \Delta(r'')} \,.
\eeq
We can then approximate $I_2$ as
\beq
I_2 = -\frac{\beta}{2r} + \frac{\beta}{k_F} \int_0^\infty \frac{dr'}{r'}\Delta(r')\,e^{-\frac{2}{k_F}\int_0^{r'} dr'' \, \Delta(r'')} \,,
\eeq
since we only make an exponentially small error in extending the integral (which is further suppressed by a factor of $k_F$) over the entire range. Thus, we approximate Eq.~\eqref{approx} as
\beq
I = -E r -\frac{\beta}{2r} + \int_0^\infty dr' \, \left(E + \frac{\beta}{k_F} \frac{\Delta(r')}{r'} \right)e^{-\frac{2}{k_F}\int_0^{r'} dr'' \, \Delta(r'')} \,.
\eeq
In the vicinity of $r^*$, we hence find that the function $\psi(r)$ (Eq.~\eqref{psi}) is approximately
\beq
\psi(r^*) \sim \frac{E r^*}{k_F} + \frac{\beta}{2 k_F r^*} - \frac{1}{k_F} \int_0^\infty dr' \, \left(E + \frac{\beta}{k_F} \frac{\Delta(r')}{r'} \right)e^{-\frac{2}{k_F}\int_0^{r'} dr'' \, \Delta(r'')} \,.
\eeq
Comparing this asymptotic behavior with Eq.~\eqref{psiapprox}, we find the vortex core energies
\beq
E = -\omega_0\,k\,l + \left(n + \frac{k-1}{2} \right)\tilde{\omega} \,,
\eeq
where
\beq
\omega_0 = \frac{1}{k_F}\frac{\int_0^\infty dr' \, \frac{\Delta(r')}{r'} e^{-\frac{2}{k_F} \int_0^{r'}dr''\,\Delta(r'')} }{\int_0^\infty dr' \, e^{-\frac{2}{k_F} \int_0^{r'}dr''\,\Delta(r'')}} \quad \text{and} \quad \tilde{\omega} = \frac{\pi}{2}\frac{k_F}{\int_0^\infty dr' \, e^{-\frac{2}{k_F} \int_0^{r'}dr''\,\Delta(r'')}} \,.
\eeq

For an elementary vortex $k=1$, $l$ is a half integer, and thus with $n = 0$, we reproduce the CdGM solution~\cite{Caroli1964}. For an MQV with $k = 2$, we see that we cannot get a zero-energy solution since $n\in \mathbb{Z}$. Moreover, for any $k$, we find $|k|$ branches of vortex core states by taking the appropriate values of $n$ such that $E \ll \Delta_0$ since our calculation is only valid in this regime. While our method does not reproduce the detailed structure of the sub-gap states away from zero energy, it allows us to analytically estimate the spectral asymmetry, since we can extract the separation between the branches from the spectrum found above.

In order to better understand the nature of the vortex core states, we now consider the specific pairing profile used in our numerical analysis,
\beq
\Delta(r) = \Delta_0 \tanh\left(\frac{r}{\xi}\right) \,.
\eeq
Since the coherence length $\xi = \frac{k_F}{\Delta_0}$, we find that
\beq
E = - \left(a \frac{\Delta_0}{k_F \xi}\right)\,k\,l +  \left(n + \frac{k-1}{2}\right)\frac{b \,k_F}{\xi} \,,
\eeq
where $a = 7 \zeta(3)/\pi^2$ and $b = \pi/2$. From this, we then see that the mini-gap $\omega_0 \sim \frac{\Delta_0}{k_F \xi}$ and furthermore, we find (pseudo) zero-energy states when
\beq
l_n = \frac{k_F \xi}{k} \left(n + \frac{k-1}{2}\right)\gamma \,,
\eeq
where $\gamma = b/a \approx 1.8$. Imposing the condition $|l|\lesssim k_F \xi$, we see that we should restrict to $n = 0, -1,-2,\dots, 1-k$ which gives us exactly $k$ branches of vortex core states. Our calculation thus demonstrates that the angular momenta where the branches cross zero energy, called crossing points in the main text, are separated by an amount $\sim k_F \xi$, in agreement with our numerical results. Furthermore, we observe that taking a different form of the pair profile simply changes the constant $\gamma$ (e.g., for $\Delta(r) = \Delta_0 \theta(r - \xi)$, we find $\gamma = 4.5$) but does not affect the scaling of the crossing points $l_n$ with $k_F$ and $\xi$. 

We note that while the spectrum for chiral states with singly quantized vortices has previously been calculated \cite{Moller2011}, the method presented here easily generalizes to chiral states with MQVs. 


\subsection{D. Observables from BdG solutions}
The particle number $\hat N$ and OAM $\hat L_z$ in the BCS ground state can be found by numerically diagonalizing the BdG Hamiltonian $H^{(l)}$. To obtain a finite spectrum we introduce a cutoff $M\gg 1$ on the radial quantum numbers such that $H^{(l)}$ is a $2M\times 2M$ Hermitian matrix. The eigenstates $(u,v)^T$ of the BdG Hamiltonian satisfy
\beq
\sum_{n'=1}^M H^{(l)}_{n,n'} \left(\begin{array}{c}
u_{n'm}^{(l)}\\
v_{n'm}^{(l)}
\end{array}
\right) = E_m^{(l)} \left(\begin{array}{c}
u_{nm}^{(l)}\\
v_{nm}^{(l)}
\end{array}
\right),
\eeq
and are normalized as $\sum_{n=1}^{M}\left( |u^{(l)}_{nm}|^2+|v^{(l)}_{nm}|^2\right)=1$. Given these, we obtain
\beq
\begin{split}
&\hat N=\sum_{l,n,m} |v^{(l)}_{nm}|^2, \\
&\hat L_z=-\sum_{l,n,m} l |v^{(l)}_{nm}|^2,
\end{split}
\eeq
where the sum over $m$ is restricted to the positive part of the energy spectrum.



\newpage 



\end{document}